\documentstyle[12pt,aps,amssymb,pre,psfig]{revtex}
\newcommand{\Z}{\mathbb{Z}}
\begin{document}

\author{I.V. Barashenkov\footnote{Email:
igor@cenerentola.mth.uct.ac.za},
V. S. Shchesnovich\footnote{Email: valery@maths.uct.ac.za}
and R. Adams\footnote{Email: rory@casimir.mth.uct.ac.za}}

\address{Department of Mathematics, University of Cape Town,
 Rondebosch 7701, South Africa}

\title{Noncoaxial multivortices
in  the complex sine-Gordon theory
on the plane}

\maketitle
\date{\today}

\vspace{10mm}
\begin{center}
{\large Abstract}
\end{center}
\begin{abstract}
We construct explicit multivortex solutions for the complex 
sine-Gordon equation (the Lund-Regge model) in two Euclidean 
dimensions. Unlike the previously found (coaxial) multivortices, 
the new solutions comprise $n$ single vortices placed at 
arbitrary positions (but confined within a finite part of the 
plane.) All multivortices, including the single vortex, have an 
infinite number of parameters. We also show that, in contrast 
to the coaxial complex sine-Gordon multivortices, the 
axially-symmetric  solutions of the Ginzburg-Landau model 
(the stationary Gross-Pitaevskii equation) {\it do not\/} 
belong to a broader family of noncoaxial multivortex 
configurations.
\end{abstract}

\vspace{10mm}


\vspace{1cm}

\newpage

\section{Introduction}
Topological solutions of nonlinear partial differential
equations on the plane  have been a subject
of intensive investigations in recent years, with applications ranging
from nonlinear optics to cosmic strings. The simplest type of topological
solitons
arise in systems involving one complex field. In this  case the solitons
realise
a $S^1 \to S^1$ map of the circle of a sufficiently large
radius on the $(x,y)$-plane into a unit circle in the
internal space $({\rm Re\/} \psi, {\rm Im\/} \psi)$
and can be classified
according to the Brouwer degree of the map, or the
winding number,
\begin{equation}
Q= \lim_{R \to \infty}
\frac{1}{2 \pi} \oint_{C_R} {\rm d} \, ({\rm arg} \psi)
= \frac{1}{2 \pi} \int_{R^2}
\epsilon_{nk} \partial_n \partial_k \, ({\rm arg}  \psi) \,  d^2x.
\label{topo_charge}
\end{equation}
One celebrated model of this kind is the Ginzburg-Landau
model (also known as the stationary  Gross-Pitaevskii equation)
arising in the description of boson condensates (in
particular, superconductors
 and superfluids) \cite{Pitaevskii,Ginzburg-Landau}:
\begin{equation}
\nabla^2 \psi + \psi (1 - |\psi|^2 )=0.
\label{GP}
\end{equation}
 (Here $\nabla =  {\bf i} \partial_x + {\bf j\/} \partial_y$.)
Another system of a similar type which received considerable
attention in  literature, is the Heisenberg ferromagnet
with  easy-plane anisotropy \cite{easy_plane}. In terms of
the stereographically projected field this can be written as
\begin{equation}
\nabla^2 \psi -
 2\frac{ (\nabla \psi)^2 \, \overline{\psi}}{1+|\psi|^2}
 + \psi \frac{1-|\psi|^2}{1+|\psi|^2}=0.
 \label{easy}
\end{equation}

In physics literature the winding number (\ref{topo_charge}) is usually
referred to as vorticity and the
planar topological solitons  as vortices.
Both the Ginzburg-Landau and the easy-plane ferromagnet equation are
well known to possess axially-symmetric solutions describing $n$
vortices sitting on top of each other, $\psi^{(n)}= e^{in \theta} \Phi_n(r)$,
where the function $\Phi_n(r) \rightarrow 1$
as $r \rightarrow \infty$.
(Here $r$ and $\theta$ are the polar coordinates on the $(x,y)$-plane.)
 Of fundamental importance,
both physically and mathematically, is
the question of whether {\it noncoaxial\/} multivortices exist.
However, despite
some encouraging recent insights \cite{OS8},  very little is known about this.

The present paper is devoted to another physically meaningful
 system\footnote{Some references to the applications
 of the complex sine-Gordon theories can be found in the concluding
 section below.} which shares a lot
of similarities with (\ref{GP}) and (\ref{easy}), the so-called
complex sine-Gordon equation:
\begin{equation}
 \nabla^2 \psi +
 \frac{ (\nabla \psi)^2 \, \overline{\psi}}{1-|\psi|^2}
 + \psi(1-|\psi|^2)=0.
 \label{sg_Euclidean}
 \end{equation}
The equation (\ref{sg_Euclidean}) is similar to (\ref{GP}) and (\ref{easy})
in that it is also an equation for one complex field on the
plane, and that it also possesses
  coaxial multivortex solutions.
An important difference, however, is that
this model is integrable and hence the complex
sine-Gordon multivortices are given by explicit
analytic formulas whereas their Ginzburg-Landau and
magnetic counterparts are available only numerically,
even in the single-vortex case.
The single-vortex solution of eq.(\ref{sg_Euclidean}) has the form
\begin{equation}
\psi^{(1)}(r, \theta)= \frac{I_1(r)}{I_0(r)} e^{i \theta},
\label{Q1}
\end{equation}
where $I_0$ and $I_1$
are the modified Bessel functions.
Using a purely algebraic
 recursive procedure \cite{BP} one can construct
  axially-symmetric
solutions of  arbitrarily high vorticity. For example, the $Q=2$
and $Q=3$
solutions are given by
\begin{equation}
\psi^{(2)}(r, \theta)=
\frac{I_1(r)^2 - I_0(r) I_2(r)}{I_0(r)^2-I_1(r)^2} \, e^{2 i \theta}
\label{Q2}
\end{equation}
and
\begin{equation}
\psi^{(3)}(r, \theta) =
\frac{[I_3(r)-I_1(r)][I_0^2(r)-I_1^2(r)]
+ I_1(r)[I_0(r)-I_2(r)]^2}
{[I_0(r)-I_2(r)][I_0(r)I_2(r)-2I_1^2(r)+I_0^2(r)]}
\, e^{3i\theta},
\label{Q3}
\end{equation}
respectively.

The primary objective of this paper is to show that the complex sine-Gordon
equation has an infinite-parameter
family of exact,
explicit, {\it noncoaxial\/} multivortices.
This family includes the
 axially-symmetric $n$-vortex solution which, therefore,  admits a
continuous
 into $n$ spatially separated single vortices.
We  calculate the energy (i.e. the Euclidean action) of the
noncoaxial $n$-vortex configuration and demonstrate that it does
not depend on the positions of the individual vortices. This
implies that the individual vortices making up the  $n$-vortex configuration
are non-interacting.

Having established the existence of a general $n$-vortex
 solution for equation (\ref{sg_Euclidean}), a natural question to ask is:
Do the Ginzburg-Landau model (\ref{GP})
and the easy-plane ferromagnet (\ref{easy}) have multivortex
solutions other than the axially-symmetric ones?
Confining ourselves to the case of the Ginzburg-Landau equation,
we attempt to answer this question via the analysis of the
spectra of linearised excitations of its symmetric solutions.
If the axially-symmetric solution admitted a $p$-parameter
nonsymmetric continuation, the corresponding linearised operator would have
$p$ zero eigenvalues.
We study the linearised spectra of the Ginzburg-Landau
axisymmetric solutions numerically;
the upshot of this study is that
they have only three zero modes related to obvious
symmetries of the equation and  therefore {\it do not\/}
belong to a broader family of noncoaxial multivortices.

On the contrary, each of the coaxial multivortices of the complex
sine-Gordon theory admits an infinite number of zero-frequency
excitations. This is an expected fact, of course, given
  the existence of the infinite-parameter nonsymmetric
configurations. However, as we show below,
the presence of the infinite number of zero modes in this case
can be established even without knowledge
of the nonsymmetric generalisations. In other words, the
possibility of the nonsymmetric continuation of the coaxial
multivortex could have been predicted simply on the basis of the
analysis of its linearisation.

The paper is organised as follows.
In the next section we derive the noncoaxial
$n$-vortex solution and in section 3 discuss  some
of its general properties. Section 4 deals with the
simplest special case of the new solution when it depends only
on one free parameter. We consider, in detail, the $n=1$,
$n=2$ and  $n=3$-solutions,
 and then extrapolate our conclusions to the situation of the
 general $n$.
The energies of the multivortices are
evaluated in section 5.
In section 6 we study, numerically, the
linearised excitations of axially-symmetric
multivortices of the Ginzburg-Landau equation
and compare them to their complex sine-Gordon counterparts.
 Finally,
some concluding remarks are made in section 7.

\section{The general multivortex solution}

As in \cite{BP}, we start with rewriting
the second-order equation (\ref{sg_Euclidean}) as a system
of two first order equations:
\begin{eqnarray}
 \overline{\partial} \psi^{(n-1)} + \psi^{(n)} (1-|\psi^{(n-1)}|^2)=0,
\label{a1} \\
 {\partial} \psi^{(n)} - \psi^{(n-1)}(1-|\psi^{(n)}|^2)=0.
\label{a2}
\end{eqnarray}
Here $\partial= \partial/\partial z$, $\overline{\partial}
=\partial/\partial \overline{z}$, and $z=(x+iy)/2$, $\overline z=
(x-iy)/2$.
This first-order system has a field-theoretic interpretation
of its own; it is nothing but the Euclidean version
of the massive Thirring model \cite{BG1,BG2}. Both
$\psi^{(n-1)}$ and $\psi^{(n)}$ satisfy equation (\ref{sg_Euclidean}),
hence eqs.(\ref{a1})-(\ref{a2}) can be seen as the B\"acklund
transformations relating two solutions of the complex
sine-Gordon equation.

Let $n=1$ in eqs.(\ref{a1})-(\ref{a2}).
For any $\psi^{(1)}$ eq.(\ref{a1}) is solved by $\psi^{(0)}=1$.
Letting $\psi^{(0)}=1$
in eq.(\ref{a2}), we get
\begin{equation}
 \partial \psi^{(1)} =1 - |\psi^{(1)}|^2.
\label{a3}
\end{equation}
Decomposing $\psi^{(1)}=f+ig$, the imaginary part of (\ref{a3}) yields
\[
\partial_x g = \partial_y f,
\]
whence we can define the potential ${\cal F}(x,y)$ such that $f=
\partial_x {\cal F}$
and $g=\partial_y {\cal F}$, or,
equivalently,
\begin{equation}
\psi^{(1)} = \overline{\partial} {\cal F}.
\label{vF}
\end{equation}
 The real part of (\ref{a3}) is then
\[
\nabla^2 {\cal F} + (\nabla {\cal F})^2=1.
\]
Letting ${\cal F}= \ln {\cal Z}_0$, this reduces to the (modified)
Helmholtz equation
\begin{equation}
\nabla^2 {\cal Z}_0 -{\cal Z}_0=0.
\label{a4}
\end{equation}
In polar coordinates, the general  solution of (\ref{a4}),
regular at the origin, is given by
\begin{equation}
{\cal Z}_0(r, \theta)= \sum_{m=0}^{\infty}
 I_m(r) \sigma_m(\theta),
\label{a5}
\end{equation}
where $I_m(r)$ is the modified
Bessel's function of order $m$
(of the first kind); $\sigma_m= \beta_m \cos(m \theta +
\delta_m)$,
and $\beta_m, \delta_m$ are
arbitrary real constants. (In particular, all $\beta_m$ with
$m$ greater than a certain $M$ can be
set equal to zero in which case the series
(\ref{a5}) becomes a finite sum.)
Returning to the variable $\psi^{(1)}$, we obtain
\begin{equation}
\psi^{(1)}= e^{i \theta}
\left\{
\frac{\sum I_m'(r) \sigma_m(\theta)}{\sum I_m (r) \sigma_m(\theta)}
+ \frac{i}{r} \frac{
\sum I_m(r) \partial_\theta \sigma_m(\theta)}
{\sum I_m(r) \sigma_m (\theta)} \right\},
\label{a6}
\end{equation}
where $I_m'= dI_m/dr$
and all sums run over $m=0,1,...\infty$.
We will assume that $\beta_0 \neq 0$; otherwise the above solution
is singular.
Without loss
of generality we can let $\sigma_0=\beta_0 \cos \delta_0=1$ in
(\ref{a6}) and  this
convention will be implied throughout this paper.
 At infinity, the solution (\ref{a6}) tends to $e^{i \theta}$;
 more precisely
\begin{equation}
\psi^{(1)}(r, \theta) = e^{i \theta}
\left\{1- \frac{1-i \kappa}{2r}+ \frac{\mu_1+i \nu_1}{r^2}
+O\left(\frac{1}{r^3}
\right) \right\} \quad
\mbox{as} \ r \rightarrow \infty,
\label{1_infty}
\end{equation}
where $\kappa, \mu_1$ and $\nu_1$ are functions of $\theta$:
\begin{equation}
\kappa= \frac{2 \partial_{\theta} \sum \sigma_m}
{\sum \sigma_m}, \quad
\mu_1= \frac{\sum (4m^2-1)
\sigma_m}{8 \sum
\sigma_m},
\quad
\nu_1=  -\partial_{\theta}
\mu_1 (\theta).
\label{kappa}
\end{equation}
Therefore, equation (\ref{a6}) gives a solution with infinite number of
parameters
and vorticity $Q=1$. For purposes of this paper we will
be referring to (\ref{a6}) as the ``general one-vortex solution".
Now the general solutions with $Q=2$ and all higher vorticities can
be obtained in a purely algorithmic way.
We simply use eq.(\ref{a1}) to
express  $\psi^{(n)}$ via $\psi^{(n-1)}$:
\begin{equation}
\psi^{(n)}=
 - \frac{1}{1-|\psi^{(n-1)}|^2} \, \overline{\partial} \psi^{(n-1)},
 \quad n=2,3,...
 \label{v2}
\end{equation}

This recursive procedure can be made rather efficient by introducing
auxiliary variables
\begin{equation}
{\cal Z}_k(r, \theta) = \sum_{m=0}^\infty\left(\gamma_m\xi_{k+m}
+ \overline{\gamma}_m\xi_{k-m}\right), \quad k=0,\pm1,\pm2,...,
\label{Zk}
\end{equation}
where
\begin{equation}
\xi_s(r,\theta) = I_s(r) e^{is\theta},\quad \gamma_m =\frac{\beta_m}{2} e^{i\delta_m}.
\label{xis}
\end{equation}
Note that ${\cal Z}_{-k} = \overline{\cal Z}_k$;
also note that the function ${\cal Z}_0$ has already been defined before
(see eq.(\ref{a5}).)   What makes the
variables ${\cal Z}_k$ useful is that the differential operators
$\partial$ and $\overline{\partial}$ act on
them simply as  index lowering and raising operators:
\begin{equation}
\partial {\cal Z}_k = {\cal Z}_{k-1},\quad
\overline{\partial} {\cal Z}_k = {\cal Z}_{k+1}.
\label{DZk}
\end{equation}
Indeed,
writing $\partial$ as
$e^{-i\theta} (\partial_r - \frac{i}{r}\partial_\theta)$,
 $\overline{\partial}$ as
$e^{i\theta} (\partial_r+\frac{i}{r}\partial_\theta)$
and
 using the identities \cite{Besseli}
\begin{equation}
\frac{{\rm d}I_m}{{\rm d}r} = \frac{I_{m-1}+I_{m+1}}{2},\quad
\frac{m I_m}{r} = \frac{I_{m-1}-I_{m+1}}{2},
\label{idents}
\end{equation}
one can readily verify that
\[
\partial \xi_m = e^{i(m-1)\theta}\left(\frac{{\rm d}
I_m}{{\rm d }r} +\frac{m}{r}I_m\right)=I_{m-1}e^{i(m-1)\theta}=\xi_{m-1},
\]
\[
\overline{\partial} \xi_m =
e^{i(m+1)\theta}\left(\frac{{\rm d} I_m}{{\rm d
}r}-\frac{m}{r}I_m\right)=I_{m+1}e^{i(m+1)\theta}=\xi_{m+1}.
\]
From here the  relations (\ref{DZk}) are straightforward.

Recalling that ${\cal F} = \ln {\cal Z}_0$ and using equation (\ref{vF}),
the general 1-vortex solution (\ref{a6}) can
be written simply as
\begin{equation}
\psi^{(1)}(r, \theta) = \frac{{\cal Z}_1}{{\cal Z}_0}.
\label{1vortex}
\end{equation}
Now applying  (\ref{v2})  and using (\ref{DZk}) gives
\begin{equation}
\psi^{(2)}(r, \theta) = \frac{{\cal Z}_1^2-{\cal Z}_2{\cal Z}_0}
{{\cal Z}_0^2-{\cal Z}_{-1}{\cal Z}_1},
\label{2vortex}
\end{equation}
\begin{equation}
\psi^{(3)}(r, \theta)  = \frac{{\cal Z}_3({\cal Z}_0^2-{\cal Z}_{-1}{\cal Z}_1) + {\cal Z}_{-1}{\cal Z}_2^2 + {\cal Z}_1^3 -2{\cal Z}_0{\cal Z}_1{\cal Z}_2}
{{\cal Z}_0({\cal Z}_0^2-2{\cal Z}_{-1}{\cal Z}_1-{\cal Z}_{-2}{\cal Z}_2) + {\cal Z}_{-1}^2{\cal Z}_2 + {\cal Z}_1^2{\cal Z}_{-2}},
\label{3vortex}
\end{equation}
and so on.
Setting
to zero all $\beta_m$ with $m \ge1$,
 eqs.(\ref{1vortex}), (\ref{2vortex}) and (\ref{3vortex}) reduce to
  the axially-symmetric vortex solutions (\ref{Q1}), (\ref{Q2})
  and (\ref{Q3}).

The explicit expressions of the multivortices
with $n \ge 4$ become cumbersome and we restrict ourselves
to producing only their asymptotic behaviours as $r \to \infty$:
\begin{equation}
\psi^{(n)}(r, \theta) e^{ -in \theta} =
 1 - \frac{n}{2r} + \frac{\mu_n(\theta)}{r^2}  +
i \left( \frac{n}{2r}{\kappa(\theta)} + \frac{\nu_n(\theta)}{r^2}
\right) +
O\left(\frac{1}{r^3}\right),
\label{BN}
\end{equation}
where $\mu_n$ and $\nu_n$ are defined by recurrence relations
\begin{eqnarray}
\mu_n &=& \mu_{n-2} + \frac{4\mu_{n-1}}{n-1} +
\frac{\partial_\theta \kappa}{2},\nonumber\\
\nu_n &=& \frac{n+1}{n-1}\nu_{n-1} - \frac{\partial_\theta\mu_{n-1}}{n-1}
-\frac{n\kappa}{4}
\partial_\theta \kappa;  \quad n \ge 2,
 \label{recur}
\end{eqnarray}
with $\mu_0=0$ and $\mu_1$, $\nu_1$ and $\kappa$ as in (\ref{kappa}).
Equations (\ref{BN})-(\ref{recur})  can be easily proved by
induction with the help of the B\"acklund transformation
(\ref{a1})-(\ref{a2}). The recurrence relation for $\mu_n$ can be
easily resolved yielding an explicit expression
\begin{equation}
\mu_n = n^2 \mu_1 +\frac{n(n-1)}{4}  \partial_\theta \kappa,
\quad n \ge 2.
\label{simple}
\end{equation}
(Unfortunately, there are no similar closed formulas for $\nu_n$.)

The relations (\ref{Zk}) can be seen as expansions over
the eigenfunctions of the angular momentum, with
$\beta_m$ being the coefficient of the eigenfunction associated
with the orbital quantum number $m$. Accordingly,
solutions (\ref{1vortex})-(\ref{BN}) can be interpreted
as orbital deformations of the axially-symmetric multivortices.
Below, in section IV, we will discuss several particular
orbital deformations in more detail.

\section{Some general properties}
\subsection{Regularity and convergence considerations}

It is not difficult to realise that  inequality
\begin{equation}
\sum_{m=1}^{\infty} |\beta_m| \le 1
\label{ineqbeta}
\end{equation}
is {\it sufficient\/} to ensure the regularity of the
general 1-vortex solution (\ref{1vortex}).
(It is not {\it necessary\/} though; see the next subsection.)
Indeed, due to the positivity of $I_m(r)$ for $r>0$ and the second identity in
(\ref{idents}), we have $I_0(r)>I_{2m}(r)$ and
$I_1(r)>I_{2m+1}(r)$ for all $m \ge 1$. Combining these inequalities
 with
   $I_0(r)>I_1(r)$ gives
$I_0(r)>I_m(r)$ for all $m \ge 1$ and  $r>0$.
This latter inequality, taken together with (\ref{ineqbeta}),
guarantees that  ${\cal Z}_0>0$ and hence the solution (\ref{1vortex}) is
bounded on the entire $(x,y)$-plane.

In case of infinitely many nonzero
coefficients $\beta_m$ we need to
 make sure  that the series in (\ref{ineqbeta}) converges.
 This can be accomplished
by imposing, for example, that
\begin{equation}
|\beta_m| \le q^m,
\label{addineq}
\end{equation}
with some $0<q<1$. The inequalities (\ref{ineqbeta}) and (\ref{addineq})
are also  sufficient for the
convergence of the series in the asymptotic formula (\ref{1_infty}).

Next, the 2-vortex solution
resulting from  the B\"acklund transformation (\ref{v2})
will only be regular if  the seed 1-vortex
solution is bounded by 1 in absolute
value, that is, if $|{\cal Z}_1|<{\cal Z}_0$. That the latter inequality
holds true can be easily
verified using the representation
\begin{equation}
{\cal Z}_k =
\frac{1}{2 \pi  i}\oint\limits_{|\ell|=1}
\frac{\mbox{d}\ell}{\ell^{k+1}}G(\ell)
e^{z\ell+\bar{z}/\ell},
\quad G(\ell) =
\sum_{m=0}^\infty\left(\gamma_m\ell^{-m}+\overline{\gamma}_m\ell^m\right)
\label{intZk}
\end{equation}
which arises by replacing $\xi_m$ in eq.(\ref{Zk}) by
\begin{equation}
\xi_m(r, \theta) = I_m(r)e^{im\theta} = \frac{1}{2 \pi i}\oint\limits_{|\ell|=1}
\frac{\mbox{d}\ell}{\ell^{m+1}}
\exp \left(z\ell+\frac{\bar{z}}{\ell}\right).
\label{xim}
\end{equation}
Eq.(\ref{xim}), in turn,
 follows from the integral formula
for the modified Bessel function \cite{Besseli}:
\[
I_m(r) = \frac{1}{2 \pi i}\oint\limits_{|\zeta|=1}\frac{\mbox{d}\zeta}
{\zeta^{m+1}} \exp \left\{\frac{r}{2}\left(\zeta+\frac{1}{\zeta}\right)\right\},
\]
where we only need to set $\zeta=\ell e^{i\theta}$.
(We remind the reader that $z=\frac12 re^{i \theta}$.)
To show that $|{\cal Z}_1|<{\cal Z}_0$, assume that the series $\sum |\beta_m|$
converges, with
eq.(\ref{ineqbeta}) being in place. The function
$G(e^{i\varphi})=1+\sum_{m=1}^\infty \beta_m\cos(m\varphi+\delta_m)$
is obviously positive for all $\varphi$ and
hence using equation (\ref{intZk}) with $\ell=e^{i\varphi}$ we get,
finally,
\[
|{\cal Z}_1|=\frac{1}{2\pi}\left|\,\int\limits_{0}^{2\pi}\mbox{d}\varphi
e^{-i\varphi}G(e^{i\varphi})e^{r\cos(\varphi+\theta)}\right|
<\frac{1}{2\pi}\int\limits_{0}^{2\pi}\mbox{d}\varphi
G(e^{i\varphi})e^{r\cos(\varphi+\theta)}={\cal Z}_0.
\]
Q.E.D.

\subsection{Translations of the vortices}
It is interesting to note that for certain choices of parameters
$\beta_1, \beta_2,...$ solutions $\psi^{(1)}$, $\psi^{(2)}$,
$\psi^{(3)}$ etc.,  describe pure translations  of the
corresponding coaxial  multivortices. The translations are
associated with {\it infinite\/} numbers of nonzero $\beta$'s.
Consider, for example, the translation along the $x$-axis:
\begin{equation}
\hat{x}=x - R, \quad \hat{y}=y.
\label{transf}
\end{equation}
 The transformation (\ref{transf})
can be written as one formula,
\[
\hat{r}\cos(\hat\theta+\varphi) =r\cos(\theta+\varphi)
- R\cos{\varphi},
\]
which holds for an arbitrary fixed angle $\varphi$.
Hence, picking up
$G(e^{i\varphi}) = e^{-R\cos{\varphi}}$
in the  representation  (\ref{intZk}), we obtain
\begin{equation}
{\cal Z}_k=
\frac{1}{2\pi}\int\limits_{0}^{2\pi}{\rm d}\varphi\,e^{-ik\varphi}
e^{-R\cos{\varphi}}e^{r\cos(\theta+\varphi)}=
\xi_k({\hat r}, {\hat \theta}),
\label{shift}
\end{equation}
where $\xi_k$ were defined in eq.(\ref{xis}):
$\xi_k(r, \theta)=I_k({r})e^{ik\theta}$.
According to (\ref{shift}), in the  reference frame $({\hat x}, {\hat y})$
our
functions ${\cal Z}_k$ of which the solution $\psi^{(n)}$ is to be built, have
the form
(\ref{Zk}) with all $\beta_m=0$ except $\beta_0=1$. In
other words, in the translated reference frame the solutions
$\psi^{(n)}$ constructed using $G(e^{i \varphi})= e^{-R \cos \varphi}$,
have the form of {\it coaxial\/} $n$-vortices.

 The orbital coefficients $\beta_m$ associated with  the translation
 can be found
 by expanding the
  function $G=e^{-R\cos{\varphi}}$  in the Fourier series:
\begin{equation}
e^{-R\cos{\varphi}}= \sum_{m=0}^\infty \left(\gamma_m e^{-im\varphi}
+\overline{\gamma}_m e^{im\varphi}\right),
\label{gmp}
\end{equation}
where
\begin{equation}
\gamma_0= \frac12 I_0(R); \quad
\gamma_m= (-1)^m I_m(R), \  \ m \ge 1.
\label{gmR}
\end{equation}
It is not difficult to check
 that the series (\ref{gmp}) converges. Indeed, the large-$m$
 asymptotic behaviour of
 the modified Bessel's function  is given by Horn's formula
 \cite{Besseli}:
\begin{equation}
I_m(R) = \frac{1}{\sqrt{2\pi}}\exp\left\{
m\left(1+\ln\frac{R}{2}\right)- \left(m+\frac12\right)\ln{m}\right\}
\left(1+{\cal O}\left(\frac{1}{m}\right)\right),\quad m\to\infty.
\label{Horn}
\end{equation}
Using (\ref{Horn}), one can easily verify that the coefficients
(\ref{gmR}) pass the ratio test:
\begin{equation}
\lim_{m\to\infty}\frac{\gamma_{m+1}(R)}{\gamma_m(R)}=0.
\label{ratio}
\end{equation}

We can normalise the coefficients according to our convention
that $\beta_0$ be equal to 1. This is done simply by replacing
$\gamma_m$ in eq.(\ref{gmR}) with
\begin{equation}
\gamma_0= \frac12; \quad
\gamma_m= (-1)^m  \frac{I_m(R)}{I_0(R)}, \  \ m \ge 1.
\label{gm_new}
\end{equation}
(Note that the corresponding $\beta_m$, $\beta_m=2(-1)^m I_m(R)/I_0(R)$,
{\it do not\/} satisfy the sufficient condition (\ref{ineqbeta}).
Despite that, the translated vortex is perfectly regular.)

Thus we conclude that the recursion procedure
(\ref{v2})-(\ref{Zk}) with an infinite
sequence of nonzero orbital  coefficients
$\gamma_m$ defined by eq.(\ref{gm_new}), gives rise to
the coaxial multivortex centred at the
point $x=R$, $y=0$. In the next section we will
show that choosing a {\it finite\/} number of nonzero $\gamma$'s
may also result in a shift of the vortex; however
that shift will  always be accompanied by a deformation.
On the contrary, the infinite sequence  (\ref{gm_new}) produces
a {\it pure\/} translation.

Our final remark in this section is
on
 the convergence of yet
another series:
\[
\sum_{m=0}^\infty m^k|\gamma_m(R)| <\infty.
\]
(This is a useful by-product of eq.(\ref{ratio}).)
The  fact that the above series converges for
all $k$ allows one to use the  asymptotic formula
(\ref{1_infty})  in the case of the translated
 1-vortex.  Restricting ourselves to terms of order $r^{-1}$,
 we find that the axially-symmetric
 (undeformed) vortex (\ref{Q1}) moved to  the point $(x=R, y=0)$,
 has the asymptotic behaviour
\begin{equation}
\psi^{(1)}(r, \theta) = e^{i\theta}\left\{1 - \frac{1}{2r} +\frac{iR\sin{\theta}}{r} +{\cal
O}\left(\frac{1}{r^2}\right)\right\}
\quad {\rm as}\; r\to\infty.
\label{shifted}\end{equation}

\section{One-parameter
 deformations of the axially-symmetric multivortices}
\label{one-parameter}

\subsection{The $n=1$ vortex}
\begin{figure}
  \begin{center}
  \psfig{file=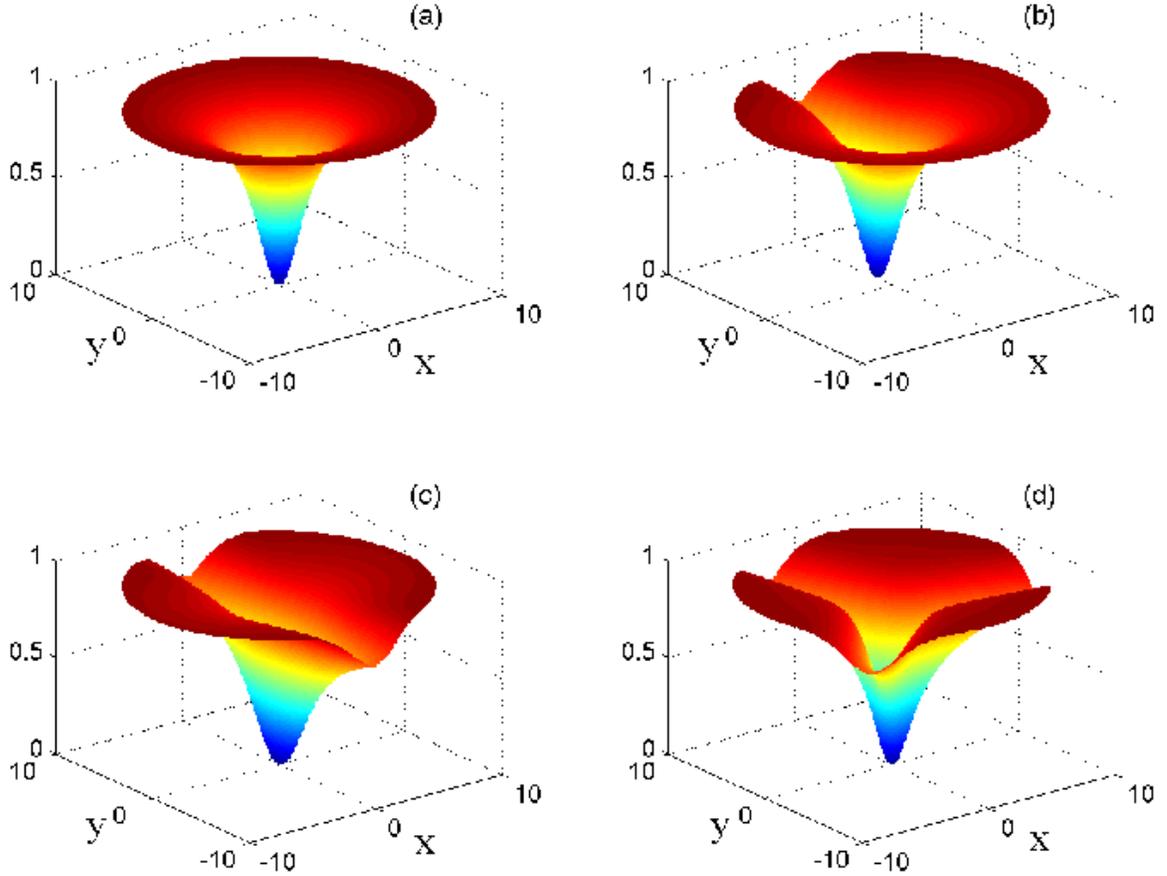,width=1.\linewidth}
  \end{center}
  \caption{\sf (a): the modulus squared $|\psi^{(1)}|^2$ of the
  unperturbed $n=1$ vortex. (b-d): its $m=1$, $m=2$ and $m=3$ orbital
  deformations. Note that in panel (b) we  used eq.(\ref{vortex1m}) with
  $\delta_1=\pi/2$  for the better visualisation.
  This gives rise to the $y$-shift of the vortex
  from the origin (and not the $x$-shift as
  in the case of  $\delta_1=0$.) Similarly, in panel (d) we set $\delta_3=\pi$.}
  \label{mods1}
  \end{figure}

In this subsection we analyse in detail the simplest situation
of a single nonzero orbital perturbation:
$\beta_k \equiv \beta \neq 0$ for $k$ equal some fixed $m$, and
$\beta_k=0$ for all other $k$. Without loss of generality we can
 set $\delta_m=0$  and consider $\beta$ to be non-negative:
 $0 \le \beta\le 1$. Eq.(\ref{1vortex}) gives
\begin{equation}
\psi^{(1)}=
\frac{I_1(r)e^{i\theta}+ (\beta/2)\left[I_{m+1}(r)e^{i(m+1)\theta}
+ I_{m-1}(r)e^{-i(m-1)\theta}\right]}{I_0(r)+\beta I_{m}(r)\cos(m\theta)}.
\label{vortex1m}
\end{equation}
The first three orbital perturbations
($m=1,2,3$) of the one-vortex solution
are shown in
figure~\ref{mods1}. The unperturbed vortex (i.e. eq.(\ref{vortex1m})
with $\beta=0$)
is  also reproduced for comparison.

The one-parameter solution (\ref{vortex1m})
is symmetric with respect to the rotation
 $\theta \to \theta + 2\pi/m$ in the $(x,y)$-plane. This
 accounts for the
number of symmetric folds in the modulus of $\psi^{(1)}$ seen in
figures~\ref{mods1}(b-d). The solution (\ref{vortex1m}) with $m>3$
is different from figures~\ref{mods1}(b-d) only in that it will
have $m$ symmetric folds.

It also follows from the cyclic symmetry that
out of all one-parameter perturbations,
only the $m=1$ perturbations  give rise to the
shift of the vortex from the origin. This shift breaks the
rotational ${\Z}_m$-symmetry completely and therefore,
is compatible only with $m=1$. This can be easily verified
by the Taylor's expansion at the origin,
\begin{equation}
\psi^{(1)}(r, \theta)= \frac{\beta_1}{2} e^{-i \delta_1}
+ {\cal O}(r).
\label{beta1}
\end{equation}
According to (\ref{beta1}), the value of
$\left. \psi^{(1)} \right|_{r=0}$ is not equal to zero
--- unless $\beta_1 =0$.

\subsection{The $n=2$ multivortex}
\begin{figure}
  \begin{center}
  \psfig{file=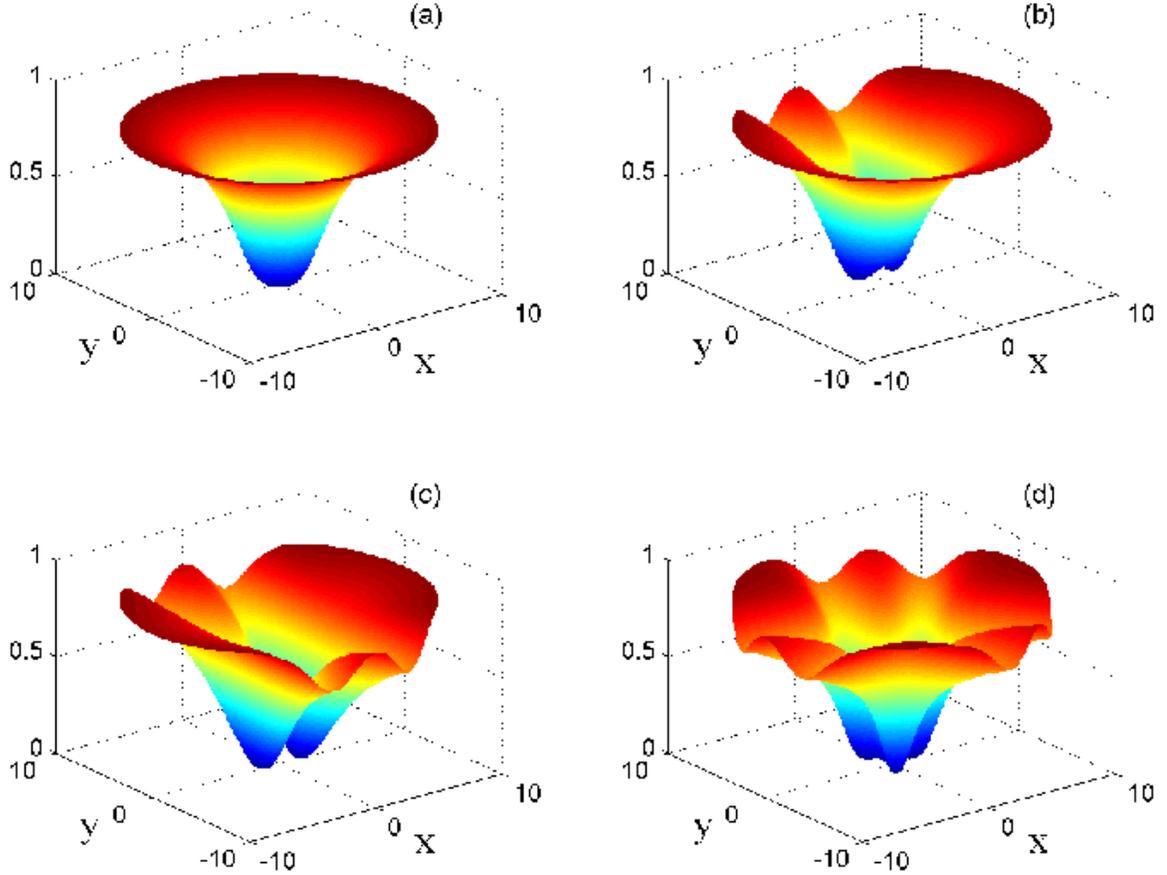,width=1.\linewidth}
  \end{center}
  \caption{\sf The modulus squared of the $n=2$-solutions.
  (a) the coaxial 2-vortex ($\beta=0$); (b) the $m=1$ perturbation
  (here $\delta_1=\pi/2$); (c) $m=2$; (d) $m=3$.
  In (b,c,d) we set  $\beta=1$ to ensure the maximum possible
  separation of individual vortices.}
  \label{mods2}
  \end{figure}

The one-parameter deformations
of the  coaxial two-vortex  with $m=1,2,3$
are shown in figure~\ref{mods2}. Each of these three perturbations
splits the repeated zero of $\psi^{(2)}(x,y)$ into several single zeros,
or, equivalently, splits the
coaxial two-vortex
 into several monovortices. The perturbations
 with $m=1$ and $m=2$ give rise to
two single vortices. For $m=2$, the indices of the newly born zeros
are equal due to the discrete rotation symmetry $\theta \to
\theta+2\pi/m =\theta+\pi$;
hence in this case the perturbed solution consists of
 two individual vortices of
vorticity $Q_1=Q_2=+1$. Here the  vorticity of the
$k$-th vortex is defined as the index of the corresponding zero of
the field:
\begin{equation}
Q_k=
\frac{1}{2 \pi} \oint_{C_k} {\rm d} \, ({\rm arg} \psi),
\label{index}
\end{equation}
where $C_k$ is a closed contour enclosing the $k$-th zero
of the solution but no other zeros.

  \begin{figure}
  \begin{center}
  \psfig{file=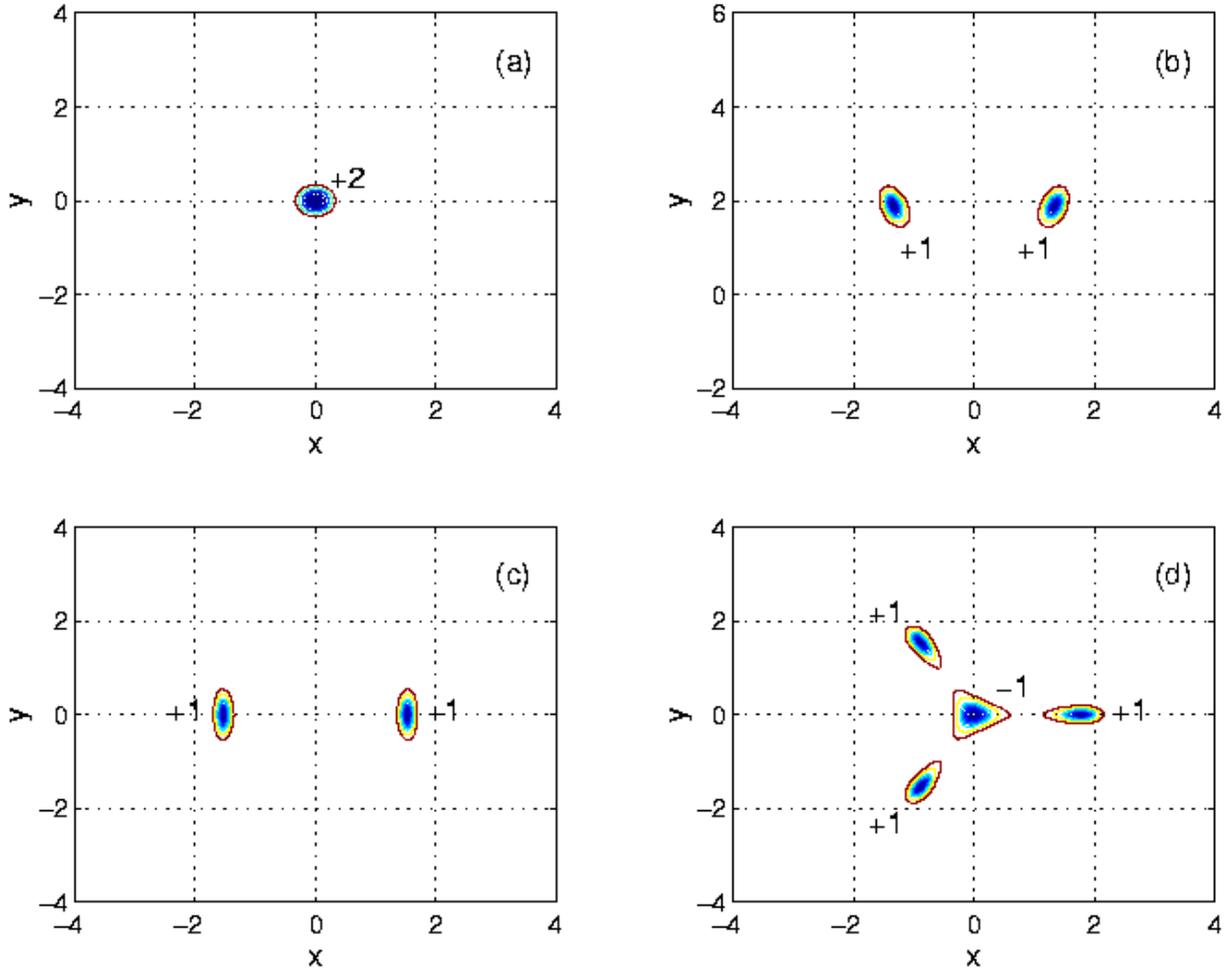,width=1.\linewidth}

  \end{center}
  \caption{\sf The level curves of $|\psi^{(2)}|^2$
  for the coaxial two-vortex (a) and its
  first three orbital perturbations:
  $m=1$  (here $\delta_1=\pi/2$) (b);  $m=2$ (c);
  $m=3$ (d). In each of these plots $\beta=1$. Only
  level curves with sufficiently low values of $|\psi|^2$ are
  shown for visual clarity.}
  \label{zeros2}
  \end{figure}

For $m=3$
there are {\it four\/} zeros.
They are not  clearly visible in figure~\ref{mods2}(d)
but become evident if one plots the level curves of
$|\psi^{(2)}|^2$, figure \ref{zeros2}.
First of all, we have
an {\it antivortex\/}
with  $Q_0=-1$ sitting at the origin.
In addition,
there are three vortices with vorticities
$Q_{1,2,3}=+1$ placed symmetrically around it.
   The fact that the origin is a zero with  index $-1$,
    can be readily concluded from the Taylor expansion for small $r$:
\begin{equation}
  \left. \phantom{\displaystyle \frac12}
  \psi^{(2)} (r, \theta) \right|_{m=3} = -\frac{\beta r}{4}e^{-i\theta} +{\cal O}(r^2).
\label{n2m3}
\end{equation}

Perturbations with higher orbital numbers do not produce
the splitting of the 2-vortex.
For all $m \ge 4$ the solution has a
 zero only at the origin. For example, for $m=4$ and $5$
the Taylor expansions about $r=0$ are:
\begin{equation}
 \left. \phantom{\displaystyle \frac12}
  \psi^{(2)} (r, \theta) \right|_{m=4} =
\frac{r^2}{8}\left(1-\frac{\beta}{2}e^{-4i\theta}\right)e^{2i\theta}
+{\cal O}(r^4),
\label{n2m4}\end{equation}
\begin{equation}
\left. \phantom{\displaystyle \frac12}
  \psi^{(2)} (r, \theta) \right|_{m=5} =
\frac{r^2}{8}e^{2i\theta} +{\cal O}(r^3).
\label{n2m5}
\end{equation}
Since $0 \le \beta \le 1$, the argument of the  term in
brackets in (\ref{n2m4}) is a periodic function of $\theta$ and
hence the index at the origin  equals +2.
The sole effect of perturbations with
higher orbital numbers on the 2-vortex solution amounts to the
symmetric deformations, similarly to the effect
of $m \ge 2$-perturbations on the 1-vortex solution.

\subsection{The $n=3$ multivortex}

\begin{figure}
\begin{center}
\psfig{file=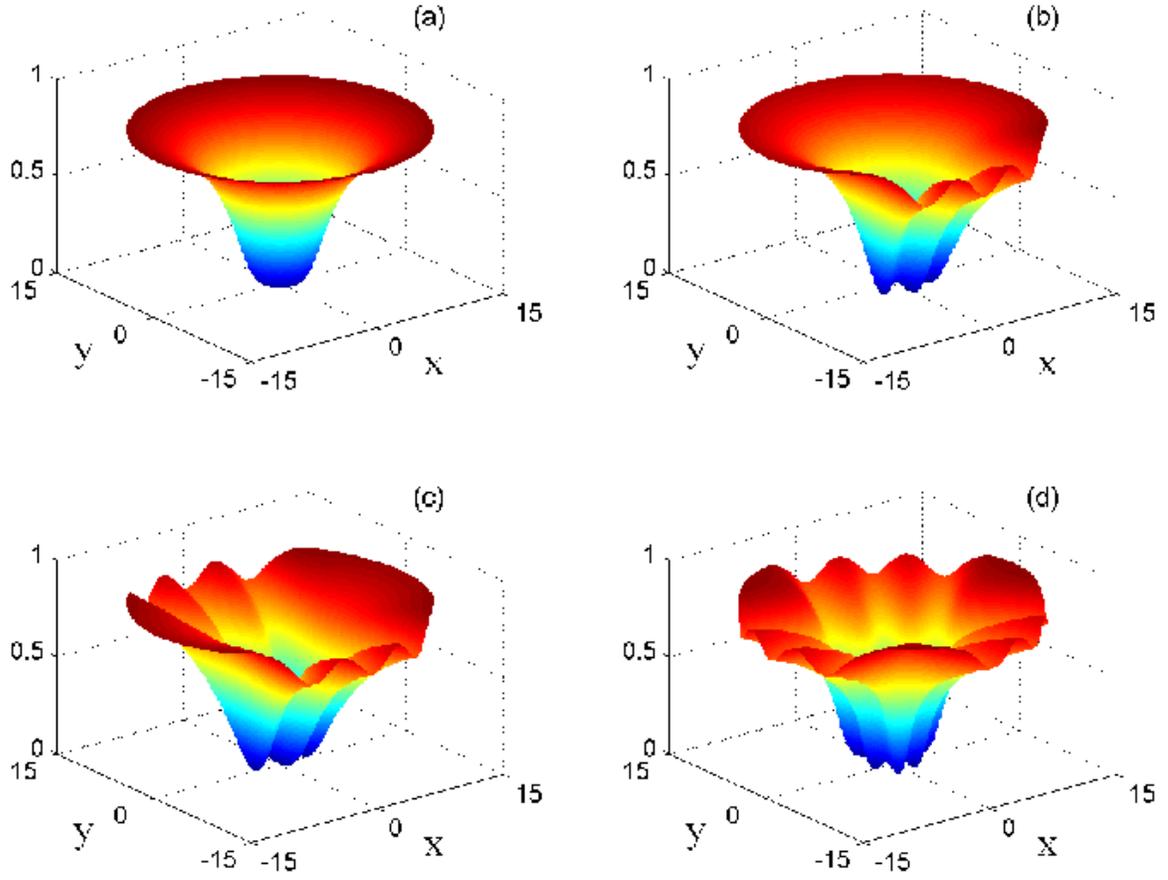,width=1.\linewidth}

\end{center}
\caption{\sf The modulus squared of the $(n=3)$-solutions.
(a) the coaxial 3-vortex ($\beta=0$); (b) its $m=1$
deformation (here $\delta_1=-\pi/2$); (c) $m=2$; (d) $m=3$.
In (b-d), $\beta=1$.}
\label{mods3}
\end{figure}

\begin{figure}
\begin{center}
\psfig{file=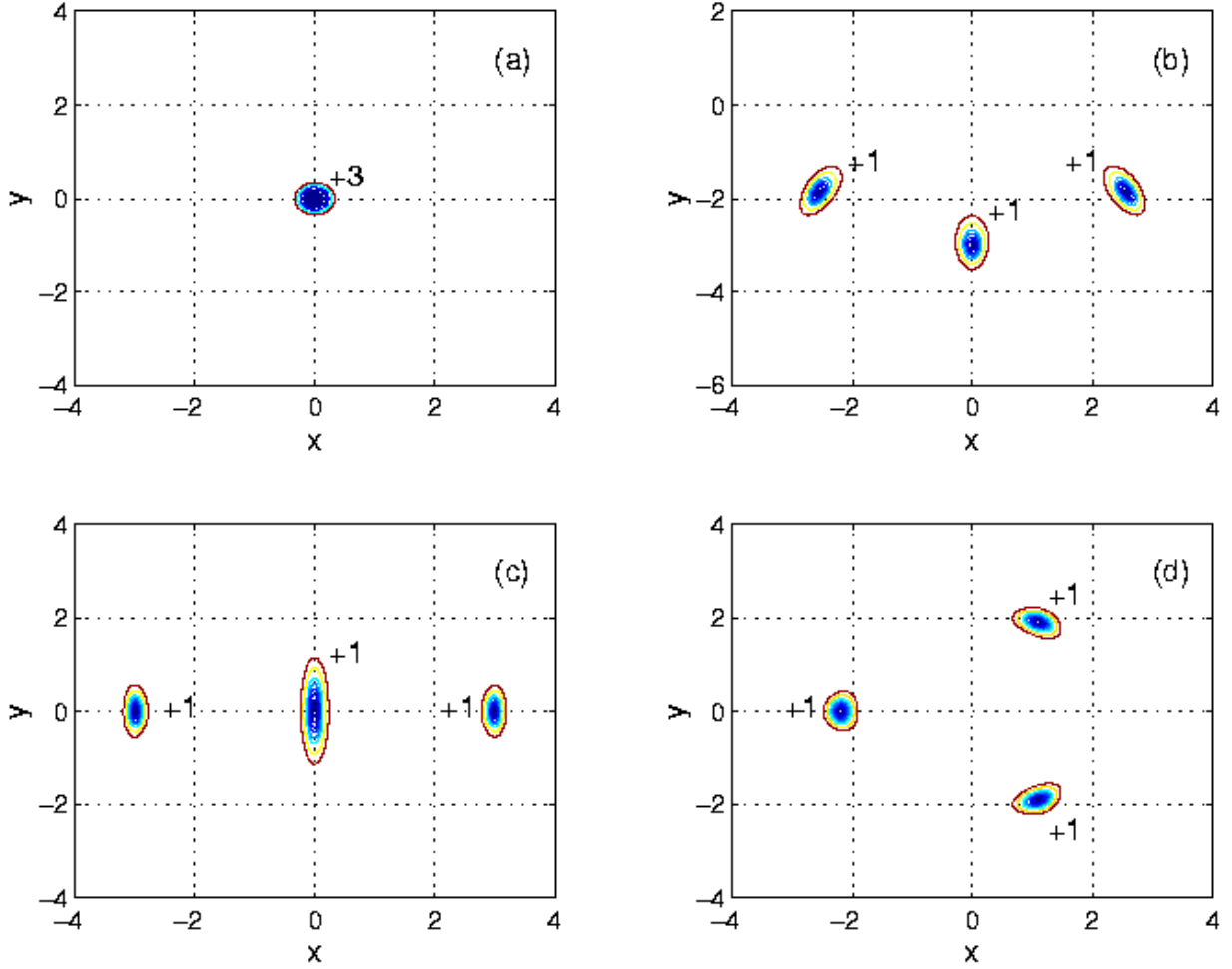,width=1.\linewidth}

\end{center}
\caption{\sf Level curves of $|\psi|^2$
for the coaxial three-vortex (a) and its lowest
 orbital deformations. (b) $m=1$  (here $\delta_1=-\pi/2$);
  (c) $m=2$; (d) $m=3$. In (b-d) $\beta=1$. }
\label{zeros3}
\end{figure}

\begin{figure}
\begin{center}
\psfig{file=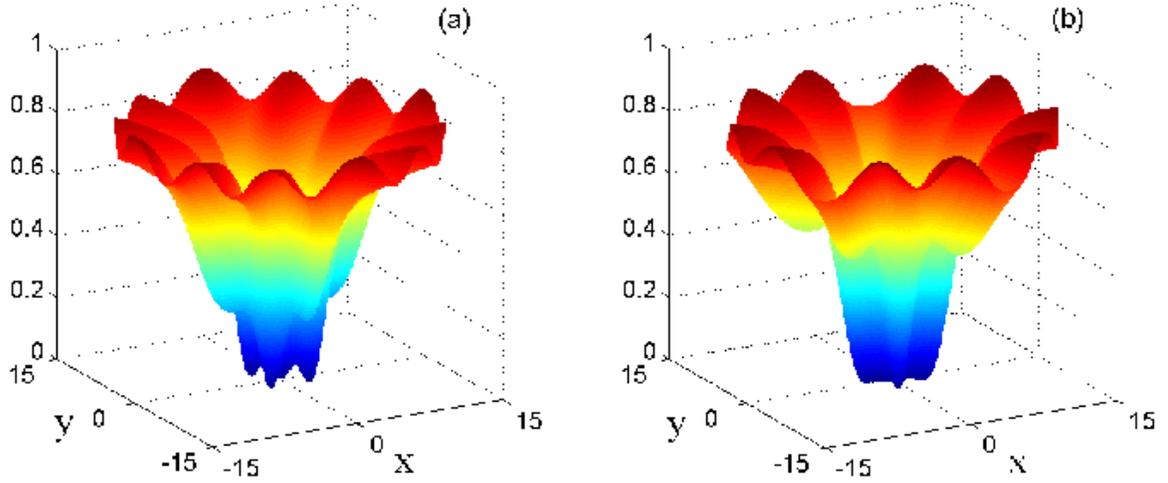,width=1.\linewidth}

\end{center}
\caption{\sf The modulus squared of the $m=4$ (a)
and $m=5$ (b) orbital deformations of the coaxial three-vortex.
Here $\beta=1$.}
\label{mods3a}
\end{figure}

\begin{figure}
\begin{center}
\psfig{file=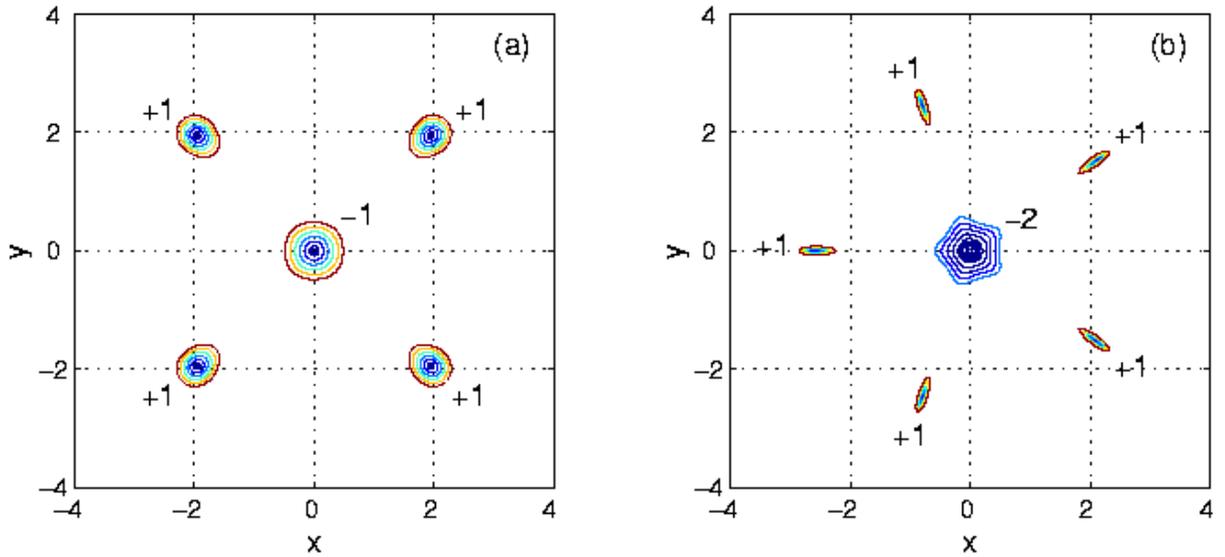,width=1.\linewidth}

\end{center}
\caption{\sf The level curves  of $|\psi|^2$ for the
two solutions shown in figure~\ref{mods3a}. (a) $m=4$; (b) $m=5$.}
\label{zeros3a}
\end{figure}

We conclude our discussions
of one-parameter deformations  with the case of the
$(n=3)$-solution. The absolute values   of $\psi^{(3)}$
are shown in figure~\ref{mods3} for $m=1,2,3$
and in figure~\ref{mods3a} for
 $m=4$ and $m=5$. (As we explain below,
 all higher orbital numbers cannot produce the splitting of the 3-vortex
  and  are less interesting, therefore.)
The location of the positions of
individual vortices is facilitated by plotting the
level curves of $|\psi^{(3)}|^2$, figures~\ref{zeros3}
and~\ref{zeros3a}.
A poor visibility of zeros in
figure~\ref{mods3a}(b) is due to a significant disproportion in
the widths of the central and surrounding vortices.
In
figure~\ref{zeros3a}(b) we have attempted to improve the
visualisation by contouring
the central vortex at lower levels of $|\psi|^2$ than the
surrounding ones.

For $m\ge2$, the indices of the surrounding zeros can be found from the
symmetry considerations provided the index of the zero at the
origin is known. The Taylor expansion about $r=0$ gives:
\begin{equation}
\left. \phantom{\displaystyle \frac12}
  \psi^{(3)} (r, \theta) \right|_{m=2} =
 -\frac{\beta}{4}\left(e^{i\theta}
+\frac{\beta/2}{4-\beta^2}e^{-i\theta}\right)r + {\cal O}(r^3),
\label{n3m2}\end{equation}
\begin{equation}
\left. \phantom{\displaystyle \frac12}
  \psi^{(3)} (r, \theta) \right|_{m=4}
   = \frac{\beta r}{4}e^{-i\theta} + {\cal O}(r^3),
\label{n3m4}\end{equation}
\begin{equation}
\left. \phantom{\displaystyle \frac12}
  \psi^{(3)} (r, \theta) \right|_{m=5} =
\frac{\beta r^2}{16}e^{-2i\theta} +{\cal O}(r^3).
\label{n3m5}\end{equation}

For higher orbital quantum
numbers, $m \ge 6$, the $(n=3)$-solution consists of just one vortex
(of vorticity $+3$)
sitting at
the origin.

\subsection{The one-parameter
deformations of the coaxial multivortices with general $n$}

Having analysed  a number of
 particular combinations of
$n$ and $m$, we can now formulate general
conjectures on the
 indices
of one-parameter
multivortex solutions.
We have identified three distinct cases depending on the
relation between $n$ and $m$.

{\bf (i.)} The simplest situation occurs if
$m \ge 2n$.
   In this case, it
is sufficient to determine the index of the vortex  at the
origin. We conjecture that
the Taylor's expansion, as $r \to 0$,  is
\begin{equation}
\psi^{(n)} =  \frac{r^n}{2^n n!} \left[1 -
(-1)^n\delta_{m,2n}\frac{\beta}{2} e^{-2in\theta}\right] \! e^{in\theta}
+ {\cal O}(r^{n+1}), \quad m\ge 2n.
\label{case1}
\end{equation}
(Here $\delta_{m,2n}$ is Kronecker's delta.)
According to (\ref{case1}), in this case we have a single multivortex 
sitting
 at the origin, with
vorticity $Q_0=+n$.

{\bf (ii.)}  Next, let $n<m<2n$.
In this case we suggest the following expansion, as $r \to 0$:
\begin{equation}
\psi^{(n)} =  \frac{\beta(-1)^{n+1} r^{m-n}}{2^{m-n+1} (m-n)!}
e^{-i(m-n)\theta} + {\cal O}(r^{n+1}), \quad
n<m<2n.
\label{case2}
\end{equation}
Eq.(\ref{case2}) shows that there is a vortex with vorticity
$Q_0=-(m-n)$ at the origin  and
$m$ vortices with indices
$Q_{1,2,...m}=+1$ placed symmetrically about the origin
in accordance with the rotation symmetry $\theta \to \theta+ 2 \pi /m$.

{\bf (iii.)} Finally, it remains to consider the case $m \le n$.
This is the most nontrivial situation as in this case there
may be no vortices at the origin at all. [See e.g.
figures~\ref{zeros2}(b),(c) and
\ref{mods3}(b),(d)].
 The analysis of several combinations of $n$ and $m$
suggests that the index $Q_0$ at the origin is the smallest-modulus
remainder  (positive or negative)
from the division of $n$ by $m$. That is, $Q_0$ is equal to  the
smallest of all positive and negative integers $q$ satisfying
\begin{equation}
n =km+q,\quad |q|<m,
\label{case3}\end{equation}
where $k$ is a positive integer.
In case there is a positive and negative remainder
of equal modulus (i.e. if $m=2l$ and
$n=k m \pm l$ with $l>0$), the index equals the
{\it positive\/} remainder: $Q_0=l$.
  Each of the $km$ surrounding vortices has  index $+1$.
  These  are placed symmetrically about
the origin (except for the case of $m=1$, of course) in
accordance with the  rotation symmetry $\theta \to \theta+ 2 \pi/m$.
 For instance,
the $n=4$, $m=2$ solution   consists of four vortices
centred on  the
$x$-axis and grouped in two pairs symmetrically with respect to the origin.

All particular solutions considered so far verify equations
(\ref{case1})-(\ref{case3}). However
 we do not yet have  proofs of these formulae in the
 case of general $n$ and $m$.

\section{The energy of the multivortices}
In this section we calculate  the
action integral for the vortices.
In the phenomenological theory of phase transitions
this integral characterises
 the
free energy of the system. (Apparently for this reason
 this quantity is also  referred to as energy
in mathematics literature \cite{Brezis,OS1,OSE}
 --- although it gives the stationary Hamiltonian only for particular
(2+1)-dimensional extensions of our (2+0)-dimensional model
(\ref{sg_Euclidean}).) Besides its fundamental
role in physical applications, the action can be used as a powerful
variational
tool in a purely mathematical analysis of stationary and moving
topological solitons (see e.g. \cite{Brezis,OS1,OSE,moduli}).

The action integral
for equation (\ref{sg_Euclidean}) has the form
\begin{equation}
E \, [\psi]= \int_{D_R} \left( \frac{|\nabla \psi|^2}{1-|\psi|^2}
+ 1- |\psi|^2 \right) d^2x
=\int_{D_R} \left( \frac{{\overline \partial} \psi \, \partial {\overline
\psi}
+ {\overline \partial} {\overline \psi} \, \partial
\psi}{2(1-|\psi|^2)}
+ 1- |\psi|^2 \right) d^2x,
\label{sg_energy}
\end{equation}
 where the integration
 is over a disc of a large radius $R$ centered
at the origin.
The massive
Thirring model (\ref{a1})-(\ref{a2}),
in its turn, is derivable from the action
\begin{equation}
S[\psi^{(n-1)},\psi^{(n)}]= \int_{D_R} \left\{
\overline {\psi^{(n-1)}}  \partial \psi^{(n)}-
\overline{\psi^{(n)}} \,{\overline \partial} \psi^{(n-1)}
+ (|\psi^{(n-1)}|^2 -1)(|\psi^{(n)}|^2 -1) + c.c.
\right\} d^2x.
\label{MTM_energy}
\end{equation}
The two systems are equivalent;
expressing $\psi^{(n-1)}$ through $\psi^{(n)}$ from eq.(\ref{a2}) and substituting
into (\ref{a1}) produces eq.(\ref{sg_Euclidean}) for $\psi^{(n)}$.
Accordingly, the integrands in
(\ref{sg_energy}) and (\ref{MTM_energy})  differ only by a divergence:
\begin{equation}
S[\psi^{(n-1)},\psi^{(n)}]=  2 E[\psi^{(n)}] -
\int_{D_R} \nabla \cdot {\bf A}^{(n)} d^2x; \quad n \ge 1,
\label{divergence}
\end{equation}
where
\begin{equation}
{\bf A}^{(n)}= - \nabla \ln (1-|\psi^{n)}|^2) +
2{\cal W}(|\psi^{(n)}|^2) \, \nabla \times  \mbox{arg} \,
\psi^{(n)},
\label{A_and_B}
\end{equation}
and
\[
{\cal W} (\rho)= \frac{\rho}{1-\rho} +\ln ({1-\rho}).
\]
 Alternatively, we can express $\psi^{(n)}$ through $\psi^{(n-1)}$;
substituting this
into (\ref{a1}) produces eq.(\ref{sg_Euclidean}) for $\psi^{(n-1)}$. This
equivalence is, again, reflected by the corresponding actions. We have
\begin{equation}
S[\psi^{(n-1)},\psi^{(n)}]=  2 E[\psi^{(n-1)}] -
\int_{D_R} \nabla \cdot {\bf B}^{(n-1)} d^2x; \quad n \ge 2,
\label{divergence2}
\end{equation}
where
\begin{equation}
{\bf B}^{(n-1)}= - \nabla \ln (1-|\psi^{(n-1)}|^2) -
2 {\cal W}(|\psi^{(n-1)}|^2) \, \nabla \times  \mbox{arg} \,
\psi^{(n-1)}.
\label{C_and_D}
\end{equation}
Both ${\bf A}^{(n)}$  and ${\bf B}^{(n-1)}$ are regular in the finite
part of the $(x,y)$-plane and therefore the double integrals in
(\ref{divergence}) and (\ref{divergence2}) can be transformed to contour
integrals over the boundary of the disc $D_R$:
\begin{equation}
S[\psi^{(n-1)},\psi^{(n)}]=  2 E[\psi^{(n)}] -
\oint_{\partial D_R}  {\bf A}^{(n)} \cdot {\hat {\bf r}} \, dl,
\quad n \ge 1,
\label{d1}
\end{equation}
\begin{equation}
S[\psi^{(n-1)},\psi^{(n)}]=  2 E[\psi^{(n-1)}] -
\oint_{\partial D_R}   {\bf B}^{(n-1)}\cdot {\hat {\bf r}} \, dl,
\quad n \ge 2.
\label{d2}
\end{equation}
Here $\hat{\bf r}= {\bf r}/r$.
For large $r$ eq.(\ref{BN}) gives
\begin{equation}
|\psi^{(n)}|^2= 1 -\frac{n}{r}
+ \frac{1}{r^2}
\left[ n^2 \left( 2 \mu_1 + \frac{1+ \kappa^2}{4}  \right)
+ \frac{n(n-1)}{2} \kappa_\theta \right]
+{\cal O}\left(\frac{1}{r^3}\right).
\label{a40}
\end{equation}
As one can easily check using (\ref{kappa}), the
${\cal O}(r^{-2})$ term in (\ref{a40}) is a total derivative:
\[
2 \mu_1+ \frac{1+ \kappa^2}{4}= - \partial^2_{\theta}
\ln \sum |\sigma_m|,
\]
and so (\ref{a40}) can be written as
\begin{equation}
|\psi^{(n)}|^2= 1 -\frac{n}{r}
+ \frac{1}{r^2} \partial_\theta \lambda^{(n)}.
\label{a21}
\end{equation}
Therefore
\begin{equation}
{\cal W}(|\psi^{(n)}|^2)= \frac{r}{n}- \ln \frac{r}{n} + \frac{\partial_\theta
\lambda^{(n)}}{n}-1 + {\cal O}\left(\frac{1}{r} \right)
\quad \mbox{as} \ r \to \infty,
\label{a22}
\end{equation}
and using this in (\ref{d1})-(\ref{d2})  we obtain, finally,
\begin{equation}
S[\psi^{(n-1)},\psi^{(n)}]=  2 E[\psi^{(n)}] -
4 \pi \left( R -n \ln \frac{R}{n} -n + \frac12 \right),
\quad n \ge 1,
\label{d3}
\end{equation}
\begin{equation}
S[\psi^{(n-1)},\psi^{(n)}]=  2 E[\psi^{(n-1)}] +
4 \pi \left( R -(n-1) \ln \frac{R}{n-1} -n + \frac12 \right),
\quad n \ge 2.
\label{d4}
\end{equation}

Subtracting (\ref{d4}) from (\ref{d3}) we arrive at the formula
relating the sine-Gordon actions of the solutions with vorticities
$n$ and $(n-1)$:
\begin{equation}
E^{(n)}-E^{(n-1)}= 2 \pi \left(
2R- n \ln \frac{R}{n} - (n-1) \ln \frac{R}{n-1} -2n+1 \right);
\quad n \ge 2.
\label{sg_diff}
\end{equation}
This equation can be used to calculate, recursively,  the energies of all
vortices; the only outstanding ingredient is the action of the
vortex with $n=1$. The $E^{(1)}$ can be found
from (\ref{d3}), provided we know the Thirring
action $S(\psi^{(0)},\psi^{(1)})$.
This action is obtained
directly from eq.(\ref{MTM_energy}) where we only need to set
$\psi^{(0)}=1$:
\begin{equation}
S(\psi^{(0)},\psi^{(1)})= \int_{D_R} ( {\overline \partial} \, \overline
{\psi^{(1)}} +\partial \psi^{(1)}) d^2x=
 \oint_{\partial D_R} \left(\overline{ \psi^{(1)}}e^{ i \theta} + \psi^{(1)}e^{ -i \theta}
 \right) \, dl.
\label{a10}
\end{equation}
Making use of the asymptotic expansion (\ref{BN}), eq.(\ref{a10}) yields
\begin{equation}
S(\psi^{(0)},\psi^{(1)})= 2 \pi (2R-1)+ {\cal O}(R^{-1}).
\label{a11}
\end{equation}
Finally, the
sine-Gordon action of the single-vortex
solution is
\begin{equation}
E^{(1)}= 2 \pi (2R-\ln R -1)+ {\cal O}(R^{-1}).
\label{a12}
\end{equation}
The formulas
(\ref{sg_diff})+(\ref{a12}) provide the actions for
vortices with any $n$.
As one could have expected, the  actions do not depend on any
of the parameters $\beta_m, \delta_m$ of the solution (\ref{a6}).
Note also
that we can use equations (\ref{sg_diff}) and (\ref{a12}) to
define the energy of the vorticity-free state $\psi^{(0)} \equiv 1$.
Consistently with one's physical intuition, eq.(\ref{sg_diff}) with $n \to 1$
yields  $E^{(0)}=0$.

Solving the recursion relation (\ref{sg_diff})
with  initial condition (\ref{a12}) we can obtain the action $E^{(n)}$
in closed form:
\begin{equation}
E^{(n)}= 2 \pi \left[
2nR-n^2 \ln R +n(\ln n-1) +
{\displaystyle 2 \sum_{k=1}^{n-1}k (\ln k-1)}
\right], \quad  n \ge 2.
\label{En_closed}
\end{equation}
Note that eq.(\ref{En_closed}) remains valid for  $n=1$ (in
which case one should simply disregard
the sum $\sum_{k=1}^{n-1}$) and for $n=0$
(in which case one should also set $n \ln n=0$).

For completeness, we also evaluate the Thirring action.
Using (\ref{En_closed}) in (\ref{d3}), we have
\begin{equation}
S[\psi^{(n-1)},\psi^{(n)}]
= 2 \pi
\left[
2(2n-1)R-2n(n-1) \ln R +
{\displaystyle 4 \sum_{k=1}^{n-1}k(\ln k-1)-1 }
\right]; \quad n \ge 2.
\label{Th}
\end{equation}
Like the previous formula, this expression remains valid for  $n=1$.

The main conclusion of this section is that the action (or ``energy")
of an $n$-vortex solution does
not depend on parameters $\beta_m$, $\delta_m$ and therefore,
on the relative positions of the
individual vortices. This implies that in any ($2+1$)-dimensional
(i.e. time-dependent) extension
of the planar complex sine-Gordon theory, vortices may form
non-interacting configurations.

\section{Complex sine-Gordon vs Ginzburg-Landau: \\
Analysis of zero modes}

The aim of this section is to find out whether the Ginzburg-Landau
axially-symmetric vortices
admit non-symmetric deformations similar to those arising in the
complex sine-Gordon equation. Let
$\psi^{(n)}$ be an axially-symmetric solution of
equation (\ref{GP}):
$\psi^{(n)}(r, \theta)=\Phi_n(r)e^{in\theta}$. If this $\psi^{(n)}$
is a member of a broader family of
solutions parameterised by $p$ continuous parameters
$\alpha_1,\alpha_2,\ldots,\alpha_p$, i.e.
\begin{equation}
    \Phi_n(r)e^{in\theta}=\psi(r,\theta;\alpha_1,\ldots,\alpha_p)\bigg|_{\alpha_1 =\alpha_2
    =\ldots= \alpha_p=0} \ ,
\end{equation}
the equation (\ref{GP}) linearised about the
symmetric multivortex will have $p$ solutions
of the form
\begin{equation}
    \delta\psi_j(r,\theta)=
    \frac{\partial\psi(r,\theta;\alpha_1,\ldots,\alpha_p)}
    {\partial\alpha_j}\Bigg|_{\alpha_1 =\alpha_2 =\ldots=
    \alpha_p=0};\quad j=1,\ldots,p.
\end{equation}
Therefore our strategy will be to examine the spectrum of
linearised perturbations of the Ginzburg-Landau vortices.
We will also be considering the
 linearisation of the complex
sine-Gordon equation (\ref{sg_Euclidean}); the
comparison of the corresponding sets of zero modes
for the two systems will give rise to some interesting
observations.

\subsection{Linearised eigenvalue problem}

Technically, it is convenient to treat the linearised boundary-value problem
as an eigenvalue problem.
With this purpose in mind, we consider the ($2+0$)-dimensional
Ginzburg-Landau equation (\ref{GP})
as a time-independent
reduction of a ($2+1$)-dimensional Higgs-field equation
\begin{equation}
    \psi_{tt}-\nabla^2\psi-(1-|\psi|^2)\psi=0.
    \label{H}
\end{equation}
(Alternatively, we could have considered it as a reduction of
the Gross-Pitaevskii equation
$i\psi_t+\nabla^2\psi+(1-|\psi|^2)\psi=0$,
but the relativistic generalisation (\ref{H}) is computationally
advantageous as it
gives rise to a symmetric eigenvalue problem.)
In a similar way we can define a ($2+1$)-dimensional
generalisation of the planar complex sine-Gordon (\ref{sg_Euclidean}):
\begin{equation}
    \psi_{tt}-\nabla^2\psi-\frac{(\nabla\psi)^2}{1-|\psi|^2}\overline\psi
    -\psi(1-|\psi|^2)=0.
    \label{csG}
\end{equation}
(Note that (\ref{csG}) is not relativistically invariant and can
hardly claim any physical relevance; we are introducing this
equation just for auxiliary purposes here.)
Assuming a solution of the form
\begin{equation}
    \psi(r,\theta,t)= \Phi_n(r) e^{in\theta} +
   \delta\psi(r,\theta,t) \equiv
    \left[ \Phi_n(r)+
    \epsilon \phi(r,\theta) \cos\omega t\right]
    e^{in\theta}
    \label{ls}
\end{equation}
and linearising (\ref{H}) in small $\epsilon$, we obtain
\begin{equation}
    -\nabla^2_r \phi-\frac{1}{r^2}(\partial_\theta+in)^2\phi-\phi+2\Phi_n^2\phi
    +\Phi_n^2\overline{\phi}=\omega^2\phi,
    \label{lH}
\end{equation}
where $\nabla^2_r\phi=\phi_{rr}+r^{-1}\phi_r$.
In a similar way, the linearisation of equation
(\ref{csG}) gives
\begin{eqnarray}
    &&-\nabla_r^2\phi  -
    \frac{1}{r^2} \partial^2_{\theta} \phi
    -\frac{2\Phi'_n\Phi_n}{1-\Phi_n^2} \,
    \partial_r \phi  -
    \frac{2in}{r^2} \frac{1}{1-\Phi_n^2}
    \partial_{\theta} \phi
    \nonumber\\
    &&+\left[
    \frac{({n^2}/{r^2})-(\Phi'_n)^2\Phi_n^2}{(1-\Phi_n^2)^2}
    +2\Phi_n^2 - 1 \right]\phi
    +\left[ \Phi_n^2
    + \frac{(n^2/r^2)\Phi_n^2-(\Phi'_n)^2}{(1-\Phi_n^2)^2}
    \right]\overline{\phi}
     =\omega^2\phi ,
    \label{lcsG}
\end{eqnarray}
where the prime over $\Phi_n$ denotes the derivative with respect to $r$.
Equations (\ref{lH}) and (\ref{lcsG}) can be
regarded as eigenvalue problems, with $\omega^2$ being an eigenvalue and
$(\phi,\overline{\phi})$ the associated eigenvector.
Expanding $\phi$ in the Fourier
series in $\theta$:
\begin{equation}
    \phi(r,\theta)=\sum^\infty_{m=-\infty}\phi_m(r)e^{im\theta}
    =\sum_{m=-\infty}^\infty\left\{
    a_m(r)+ib_m(r)\right\}e^{im\theta},
    \label{F}
\end{equation}
and transforming to
\begin{equation}
    u_m(r)=a_m+a_{-m}, \qquad v_m(r)=a_{m}-a_{-m},
    \label{EF}
\end{equation}
we obtain a sequence of one-dimensional eigenvalue problems,
one for each value of the azimuthal number
$m$:
\begin{equation}
   {\cal L}_{n,m}\left({u_{m} \atop v_{m}} \right)
    =\omega^2\left({u_{m} \atop v_{m}} \right).
    \label{EV}
\end{equation}
The operator ${\cal L}_{n,m}$ is defined by
\begin{equation}
    {\cal L}_{n,m}\equiv\left({-\nabla^2_r +
    \frac{n^2+m^2}{r^2} +3\Phi_n^2(r)-1
    \atop
    \frac{2mn}{r^2}}  {\frac{2mn}{r^2}
    \atop -\nabla^2_r + \frac{n^2+m^2}{r^2} +\Phi_n^2(r)-1}
    \right)
    \label{LGL}
\end{equation}
in the Ginzburg-Landau case, and by
\begin{equation}
    {\cal L}_{n,m}\equiv\left({-\nabla_r^2 +
    {\frak B}_n(r)\frac{{\rm d}}{{\rm d}r} + \frac{m^2}{r^2} +{\frak
    C}_n(r)+{\frak D}_n(r)
    \atop m{\frak A}_n(r)}
    {m{\frak A}_n(r) \atop -\nabla_r^2 + {\frak B}_n(r)\frac{{\rm d}}{{\rm d}r} +
    \frac{m^2}{r^2} +{\frak C}_n(r)-{\frak D}_n(r)}\right)
    \label{LsG}
\end{equation}
in the complex sine-Gordon case.
In (\ref{LsG}) we have introduced the notations
\begin{eqnarray}
    {\frak A}_n(r)&=& \frac{2n}{r^2}\frac{1}{1-\Phi_n^2},
    \nonumber\\
    {\frak B}_n(r)&=&
    \frac{2n}{r}\frac{\Phi_n^2}{1-\Phi_n^2}-2\Phi_n\Phi_{n-1},
    \nonumber\\
    {\frak C}_n(r)&=& 2\Phi_n^2 -\Phi_n^2\Phi_{n-1}^2 -1
    +\frac{n^2}{r^2}\frac{1+\Phi_n^2}{1-\Phi_n^2}
    +\frac{2n}{r}\frac{\Phi_{n-1}\Phi_n^3}{1-\Phi_n^2},
    \nonumber\\
    {\frak D}_n(r)&=& \Phi_n^2-\Phi_{n-1}^2 +
    \frac{2n}{r}\frac{\Phi_n\Phi_{n-1}}{1-\Phi_n^2}.
    \label{coe}
\end{eqnarray}
The imaginary parts of the Fourier coefficients $\phi_m$
satisfy
the same eigenvalue problem (\ref{EV}) with ${\cal L}_{n,m}$ as in
 (\ref{LGL}) or (\ref{LsG}),
 where the eigenfunctions $(u_m,v_m)$ should only be
defined by
\begin{equation}
    u_m(r)=b_{m}-b_{-m},\qquad v_m(r) = b_m+b_{-m}.
    \label{EF2}
\end{equation}
(Note that the eigenfunctions  $(u_m,v_m)$ do depend on the
vorticity $n$ but we are omitting the corresponding subscript
to keep the notation simpler.)

Thus the spectrum of linearised excitations
of the symmetric multivortex $\psi^{(n)}$ is given by
eigenvalues of the operator ${\cal L}_{n,m}$ with $m$ varying from
$-\infty$ to $\infty$.
In fact in view of equations (\ref{EF})
and (\ref{EF2}) we can restrict ourselves to nonnegative $m$ only.
If $(u_m,v_m)$ is an eigenvector
of the operator ${\cal L}_{n,m}$ associated with an eigenvalue $\omega^2$,
then $(u_m,-v_m)$ is an
eigenvector of the operator ${\cal L}_{n,-m}$ associated with the same
eigenvalue $\omega^2$.

We solved the eigenvalue problem (\ref{EV})-(\ref{LGL}) numerically.
Before discussing the results of the numerical analysis, it is
instructive to compare the spectrum structure of the Ginzburg-Landau
operator (\ref{LGL}) with
that of its complex sine-Gordon counterpart.

\subsection{The spectrum structure}
For any positive $n$ and $m \ge 0$ we  introduce two
bases of solutions of the linear system (\ref{EV})-(\ref{LGL}).
One basis can be defined by the asymptotic behaviours at the origin.
To this end, we  rewrite the system
in terms of ${\tilde u}=(u_m+v_m)/2$ and ${\tilde v}=(u_m-v_m)/2$
(where  the subscript $m$  is omitted for simplicity of notation):
\begin{equation}
\left(
\begin{array}{cc}
L_{m+n}+ 2 \Phi_n^2  & \Phi_n^2 \\
\Phi_n^2  &    L_{m-n}+ 2 \Phi_n^2
\end{array}
\right)
\left(
\begin{array}{c}
{\tilde u} \\ {\tilde v}
\end{array}
\right)
= \omega^2
\left(
\begin{array}{c}
{\tilde u} \\ {\tilde v}
\end{array}
\right).
\label{pq}
\end{equation}
Here $L_s= -\nabla^2 +(s^2/r^2)-1$.
  The advantage of the formulation
(\ref{pq}) is in that for $r \to 0$, the
cross-coupling potentials  $\Phi_n^2 \sim r^{2n}$ are small
and the equation for  ${\tilde u}(r)$
decouples from the equation for ${\tilde v}(r)$. Expanding each of the ${\tilde u}(r)$
and ${\tilde v}(r)$ in power series in $r$ (supplemented by logarithmic
terms where necessary), e.g.
\[
{\tilde u}(r)=({\tilde u}_0 + {\hat u}_0 \ln r) r^{\frak p} + ({\tilde u}_1 +
{\hat u}_1 \ln r) r^{{\frak p}+2} +...
\]
and substituting into (\ref{pq}), one can easily verify that
for each pair $(n,m)$ there are four solutions with the following
asymptotic behaviour:
\begin{eqnarray*}
{\tilde Z}_1=
\left(
\begin{array}{c}
r^{m+n}[1+o(r)] \\ {\cal O}(r^{m+n+2n+2})
\end{array}
\right); \quad
{\tilde Z}_2=
\left(
\begin{array}{c}
{\cal O}(r^{|m-n|+2n+2}) \\
r^{|m-n|}[1+o(r)]
\end{array}
\right); \quad
{\tilde Z}_3=
\left(
\begin{array}{c}
r^{-(m+n)}[1+o(r)] \\ o(r^{-(m+n)+2n+1})
\end{array}
\right);
\\
{\tilde Z}_4=
\left(
\begin{array}{c}  o(r^{-|m-n|+2n+1}) \\
r^{-|m-n|}[1+o(r)]
\end{array}
\right) \
(\mbox{for} \ m \neq n); \quad
{\tilde Z}_4=
\left(
\begin{array}{c}
{\cal O} (r^{2n+2} \ln r) \\
 \ln r \cdot [1+o(r)]
\end{array}
\right) \
(\mbox{for} \ m=n).
\end{eqnarray*}
(Here ${\tilde Z}$ denotes  the column $({\tilde u},{\tilde v})^T$, of course.)
Transforming back to $u_m$ and $v_m$
and introducing the notation $Z=(u_m,v_m)^T$, we have four linearly
independent solutions of the system (\ref{EV})-(\ref{LGL})
(as $r \to 0$):
\[
Z_1= r^{m+n}[1+o(r)]
\left(
\begin{array}{c}
1 \\ 1
\end{array}
\right);
\quad
Z_2= r^{|m-n|}[1+o(r)]
\left(
\begin{array}{r}
1 \\ -1
\end{array}
\right);  \quad
Z_3= r^{-(m+n)}[1+o(r)]
\left(
\begin{array}{c}
1 \\ 1
\end{array}
\right);
\]
\begin{eqnarray}
 Z_4  = r^{-|m-n|}[1+o(r)]
\left(
\begin{array}{r}  1 \\
-1
\end{array}
\right) \
(\mbox{for} \ m \neq n)&,& \nonumber \\
 Z_4 = \ln r \cdot [1+o(r)]
\left(
\begin{array}{r}  1 \\
-1
\end{array}
\right) \
(\mbox{for} \ m=n)&.&
\label{Y}
\end{eqnarray}
The solutions $Z_1, Z_2$ are bounded and  $Z_3, Z_4$ unbounded
for all $m$ and $n$.

In a similar way one can show that
 the linearised complex
sine-Gordon (\ref{EV})-(\ref{LsG}) also has two bounded and
two unbounded solutions near the origin. In this case one should only
take into account that as $r \to 0$,
\begin{eqnarray*}
\begin{array}{ll}
{\displaystyle
 {\frak A}  =\frac{2n}{r^2}+ \frac{2n}{(2^n n!)^2}r^{2n-2}+
{\cal O}(r^{2n}),} \ \
&
{\displaystyle
{\frak B}=-\frac{1}{r}-\frac{2n}{(2^n n!)^2}r^{2n-1} +{\cal O}(r^{2n+1}),}
\\
{\displaystyle
 {\frak C}=\frac{n^2}{r^2}-1+\frac{2n^2}{(2^n n!)^2} r^{2n-2}+
{\cal O}(r^{2n}),} \ \
&
{\displaystyle
{\frak D}={\cal O}(r^{2n})}.
\end{array}
\end{eqnarray*}
  The four basis solutions are given
by the same equations (\ref{Y}) as in the
case of the Higgs field.  We should emphasise here that eqs.(\ref{Y})
are valid for all $\omega$ (including $\omega=0$),
both in the Higgs and sine-Gordon cases.

The second basis is defined by the asymptotic behaviours as
$r \to \infty$.  Consider, first, the
Higgs system (\ref{EV})-(\ref{LGL})
and let $0 < \omega <\sqrt2$. The four solutions are given by
\begin{eqnarray}
{Y}_{1,2}=
\frac{e^{ \pm i \omega r}}{\sqrt{r}}
\left( \begin{array}{r}
-\frac{mn}{r^2} +{\cal O}\left(\frac{1}{r^3}\right)\\
1 \pm \frac{(1/4)-m^2}{2i \omega r}
+{\cal O}\left(\frac{1}{r^2}\right)
\end{array} \right);
\label{Yt12} \\
{Y}_{3,4}=
\frac{e^{ \pm \sqrt{2-\omega^2} r}}{\sqrt{r}}
\left( \begin{array}{r}
1 \pm \frac{2n^2+(1/4)-m^2}{2 \sqrt{2-\omega^2} r}  + {\cal O}\left(
\frac{1}{r^2}\right)  \\  \frac{mn}{r^2} +{\cal O}\left(\frac{1}{r^3}\right)
\end{array} \right).
\label{Yt34}
\end{eqnarray}
The solutions ${Y}_{1}$, ${Y}_{2}$,   and ${Y}_{4}$
are bounded and ${Y}_{3}$ unbounded as $r \to \infty$.
Similarly,   for all $0<\omega<2$
the linearised complex sine-Gordon (\ref{EV})-(\ref{LsG})
 has three bounded and one unbounded solution
 as $r\to\infty$:
\begin{eqnarray}
{Y}_{1,2}=  \frac{e^{ \pm i \omega r}}{r}
\left( \begin{array}{r}
-\frac{m}{2r}
  +{\cal O}\left(\frac{1}{r^2}\right)\\
 1+  \left(\frac{1-2m^2}{16} \pm \frac{m^2n}{4i \omega} \right)
\frac{1}{r^2}+  {\cal O}\left(\frac{1}{r^3}\right)
\end{array} \right);
\label{Ycs12} \\
Y_{3,4}=
e^{ \pm \sqrt{4-\omega^2} r} r^{\frak p}
\left( \begin{array}{r}
1 \mp \frac{2m^2-1-{\frak p}({\frak p}+1)}
{2 \sqrt{4-\omega^2}r} +{\cal O}\left(\frac{1}{r^2}\right)
  \\  \frac{m}{2r} +{\cal O}\left(\frac{1}{r^2}\right)
\end{array} \right),
\label{Ycs34}
\end{eqnarray}
where ${\frak p}=\mp 2n(4-\omega^2)^{-1/2} -1$.
In the derivation of (\ref{Ycs12})-(\ref{Ycs34}) we made use of the
asymptotic expansions, as $r \to \infty$,
 of the coefficient functions in (\ref{coe}):
\begin{eqnarray}
    {\frak A}_n(r)&=& \frac{2}{r} -\frac{1}{4r^3}-\frac{n}{2r^4}+
{\cal O} \left(\frac{1}{r^5}\right),
    \nonumber\\
    {\frak B}_n(r)&=&
   - \frac{2}{r}-\frac{1}{4r^3}+{\cal O} \left(\frac{1}{r^4}\right),
    \nonumber\\
    {\frak C}_n(r)&=& 2-\frac{2n}{r}- \frac{1}{2r^2}
-\frac{3n}{4r^3} +{\cal O} \left(\frac{1}{r^4}\right),
 \nonumber\\
    {\frak D}_n(r)&=& 2-\frac{2n}{r}- \frac{1}{2r^2}
-\frac{3n}{4r^3} +{\cal O} \left(\frac{1}{r^4}\right).
\label{coexp}
\end{eqnarray}

Each of the solutions $Z_i$ ($i=1,...4$) can be expanded over the
basis ${Y}_j$:
\[
Z_i(r)= \sum_{j=1}^4 T^{(n,m)}_{ij}(\omega)  {Y}_j(r).
\]
For all $0<\omega<\sqrt2$ in the Higgs case and all  $0<\omega <2$ in the
complex sine-Gordon case,
the linear combination $T^{(n,m)}_{23}Z_1(r)-T^{(n,m)}_{13}Z_2(r)$
represents a solution which is bounded both as $r \to 0$ and $r \to \infty$.
Therefore, in both cases small nonzero $\omega$
belong to the continuous spectrum. The question that is of concern to us,
of course, is whether $\omega=0$ belongs to the continuum;
in other words, is there a bounded solution for $\omega=0$?
Surprisingly, the answers for the Higgs (alias Ginzburg-Landau)
 and complex
sine-Gordon linearisation, are different.

Let us start with the Higgs case and let $r \to \infty$. Two asymptotic
solutions, ${Y}_{3,4}$, are given by eq.(\ref{Yt34}) where
we only need to set  $\omega=0$.
The other two asymptotic formulas,
eq.(\ref{Yt12}), cannot be used for $\omega=0$.
Instead, we have two
solutions with the asymptotics
\begin{eqnarray}
{Y}_{1,2}=
r^{\pm m}
\left( \begin{array}{r}
-\frac{mn}{r^2} +{\cal O}\left(\frac{1}{r^4}\right)\\
 1
+{\cal O}\left(\frac{1}{r^2}\right)
\end{array} \right) \ (\mbox{for} \, m \neq 0);
\nonumber \\
Y_1=
{\textstyle \ln r \cdot
\left[1+ {\cal O}\left( \frac{1}{r^2} \right)\right]
}
\left( \begin{array}{c}
0\\
 1
\end{array} \right);
\quad
Y_2=
{\textstyle
\left[1+ {\cal O}\left( \frac{1}{r^2} \right)\right]
}
\left( \begin{array}{c}
0\\
 1
\end{array} \right) \ (\mbox{for} \, m = 0).
\label{Yt0}
\end{eqnarray}

Therefore, for $\omega=0$ we only have two, not three, solutions
bounded as $r \to \infty$: ${Y}_2$ and  ${Y}_4$.
The solutions $Z_1$ and $Z_2$, bounded at the origin, can still
be expanded over the basis ${Y}_j$ $(j=1,...4)$; however,
this time
in order for the linear combination $c_1 Z_1+ c_2 Z_2$ to
remain bounded as $r \to \infty$, the constants $c_1$
and $c_2$ have to satisfy {\it two\/} conditions:
$c_1 T^{(n,m)}_{11}(0)+c_2 T^{(n,m)}_{21}(0)=0$
and
$c_1 T^{(n,m)}_{13}(0) +c_2 T^{(n,m)}_{23}(0)=0$.
This imposes a requirement on
the matrix $T^{(n,m)}(\omega)$ at the point $\omega=0$:
\begin{equation}
\Delta^{(n,m)} \equiv
\left|
\begin{array}{cc}
T^{(n,m)}_{11}(0) & T^{(n,m)}_{21}(0) \\
T^{(n,m)}_{13}(0) & T^{(n,m)}_{23}(0)
\end{array}
\right|
=0,
\end{equation}
which is not  {\it a priori\/} satisfied for all values of $n$ and $m$.
Consequently, the zero mode (i.e. a solution $u_m(r),v_m(r)$ pertaining to
$\omega=0$ and bounded for all $0 \le r < \infty$) can only arise
 for {\it some\/} pairs $(n,m)$.

Consider now the linearised complex sine-Gordon (\ref{EV})+(\ref{LsG}).
As in the Higgs case, the asymptotic solutions  ${Y}_{3,4}(r)$
(eqs.(\ref{Ycs34})) are valid for $\omega=0$ whereas the
 formulas
  (\ref{Ycs12}) are not. Instead,
for $\omega = 0$ the two
solutions ${Y}_{1,2}(r)$ have the asymptotics
\begin{equation}
{Y}_1= \frac{1}{r}
\left( \begin{array}{r}
-\frac{m}{2r} +{\cal O}\left(\frac{1}{r^2}\right)\\
 1+{\cal O}\left(\frac{1}{r^2}\right)
\end{array} \right); \quad
{Y}_2=
\left( \begin{array}{r}
-\frac{m}{2r}
 +{\cal O}\left(\frac{\ln r}{r^2}\right)\\
 1+m^2 n \frac{\ln r}{r}
+ {\cal O}\left(\frac{\ln r}{r^2}\right)
\end{array} \right),
\quad r \to \infty.
\label{Ycs02}
\end{equation}
Thus in the complex sine-Gordon case we have {\it three\/}
bounded solutions as $r \to \infty$:  ${Y}_{1}$,
${Y}_2$, and ${Y}_4$
(and not two as in the Higgs case.) Accordingly,
there always is a
solution bounded for all $r \in [0, \infty)$:
 $(u_m,v_m)=T^{(n,m)}_{23}Z_1(r)-T^{(n,m)}_{13}Z_2(r)$, and so
  we may conclude that
there is a zero mode for any pair $(n,m)$ --- without
having to find this solution analytically or numerically.
In this sense, in the complex sine-Gordon case the value
$\omega=0$ belongs to the continuous spectrum whereas
in the Ginzburg-Landau case
  zero modes can only
appear as discrete eigenvalues. In the
latter case we had to resort to the help
of computer.

\subsection{Ginzburg-Landau zero modes:
a numerical search}
To find eigenvalues and eigenfunctions numerically, we replaced derivatives
in
eq.(\ref{LGL}) (and
 eqs.(\ref{org})-(\ref{inf}) below)
 with the second order-accurate finite differences. Typically we took a
grid with $1000$ points on an interval $0<r\le80$ (i.e. with the step size
$\Delta r=0.08$). The
coaxial multivortices $\Phi_n(r)$ of which
the potentials in (\ref{LGL}) are formed, were
pre-computed using Newton's method.

The appropriate
boundary conditions at the origin follow from the asymptotic
behaviours (\ref{Y}):
\begin{eqnarray}
   u_r(0)=v_r(0)=0, \quad \mbox{for} \ m \neq n \pm 1;
\nonumber \\
u(0)=v(0)=0, \quad \mbox{for} \ m = n \pm 1.
\label{org}
\end{eqnarray}
At infinity, the boundary conditions are chosen to accomodate the
bounded solutions in
eqs.(\ref{Yt34})+(\ref{Yt0}):
\begin{equation}
 u_r\to0,\ v_r\to0\quad  \mbox{as} \ r\to\infty.
    \label{inf}
\end{equation}

Numerically, zero modes appear as small nonzero eigenvalues
and one still has to distinguish them from   small eigenvalues
arising from the continuous spectrum when the infinite line is
approximated by a finite interval $[0,80]$. The
genuine zero modes can be discerned by considering the
large-$r$ behaviours of the associated eigenvectors.
The ``true" zero modes are allowed to decay
exponentially (as in (\ref{Yt34})) or as $r^{-m}$
(as in (\ref{Yt0})) whereas the continuous spectrum solutions
will generically decay as $1/\sqrt{r}$ (see eq.(\ref{Yt12})).

Before calculating the eigenvalues of the Higgs vortices, we tested
our numerical scheme on the complex sine-Gordon operator
(\ref{LsG}). As expected,
 we obtained one zero mode for each $n$ and $m$. (We tested
$m=0,1,...,5$ for each of $n=1,2,3$.) The
associated eigenfunctions were found to coincide with the
derivatives of the general $n$-vortex solution $\psi^{(n)}$
w.r.t. the azimuthal parameters $\beta_m$ (see the next subsection.)

Proceeding to the Higgs system,
we examined the one-, two and three-vortex solutions,
i.e., $n=1,2$ and $3$.
The azimuthal quantum number of
the analysed perturbation ranged from $m=0$ to $5$ in each case.
Zero eigenvalues were only found
for $m=0$ and $1$; both result from obvious symmetry properties
of the Ginzburg-Landau/Higgs equation.

For $m=0$, the complex eigenfunction $\phi(r)$
 of the
operator (\ref{lH}) associated with the numerically-found zero eigenvalue,
was found to coincide
with the function $i\Phi_n(r)$ (i.e., $u_0=0$, $v_0=2 \Phi_n$.)
 This zero mode is related to the $U(1)$
invariance of the
Ginzburg-Landau equation (\ref{GP}):
\begin{equation}
    \phi(r)e^{in\theta}=\frac{\partial}{\partial\alpha}
    \Phi_n(r)e^{in\theta+i\alpha}\bigg|_{\alpha=0}.
\end{equation}

For $m=1$, the numerical
eigenvector $(u_1,v_1)$ associated with the zero eigenvalue
was found to be equal to
$(\Phi_n', - \frac{n}{r} \Phi_n)$.
(There is, of course,
  a zero mode for $m=-1$ as well: $(u_{-1},v_{-1})=
(\Phi_n',  \frac{n}{r} \Phi_n)$.)
The corresponding complex perturbations (\ref{ls})
are the translation modes,
one $\phi e^{in \theta}$ given by
$\partial_x\left[\Phi_n(r)e^{in\theta}\right]$
and the other one by $\partial_y\left[\Phi_n(r)e^{in\theta}\right]$.

Both the translational and the $U(1)$-zero modes
 are well known to workers in this field. (See e.g.
\cite{Pitaevskii,Hagan,OS1}).  There is a simple analytical argument
showing that zero modes cannot arise for $m \ge 2n$
 \cite{OS1}; this fact is also known to  specialists.
However, the nonexistence of
zero modes for $1<m< 2n$ does not seem to have appeared in literature
before.

Since the only zero-frequency excitations of the axially-symmetric vortices
are those associated with the phase shifts and translations, we
conclude that the coaxial multivortices of the stationary
Ginzburg-Landau model {\it do not\/}
admit continuous nonsymmetric deformations. In particular, two vortices
sitting, symmetrically, on
top of each other cannot be continuously separated. This does not mean,
of course, that the
Ginzburg-Landau model does not admit noncoaxial multivortex solutions at all.
Multivortex
configurations with {\it finite\/} separations between individual vortices
may
exist (and in
fact there are indications that they do exist \cite{OS8}). However, the
intervortex separations
will only admit discrete sets of values, or will be allowed to vary
continuously but be bounded
from below by certain finite distances.

This is in sharp contrast with the coaxial multivortices of the
complex sine-Gordon theory which
can be continuously split and moved apart.
As we have shown in previous sections, each
axially-symmetric solution of this model is a member of an infinite-parameter family of solutions
corresponding to a specific choice of azimuthal deformation parameters:
$\beta_1=\beta_2=\ldots=0$.
 We conclude this section by demonstrating that each of these
continuous parameters gives rise to a zero mode in the spectrum of the corresponding linearised
operator (\ref{LsG}).

\subsection{Zero modes of the complex sine-Gordon multivortices}
The derivative of the general $n$-vortex solution w.r.t. the azimuthal
parameter $\beta_m$ ($m \ge 1$), is given by
\begin{equation}
\left. \frac{\partial}{\partial \beta_m} \psi^{(n)}
\right|_{\beta_1=\beta_2=...=0}=\left.
\frac12 \sum_{k=1-n}^n
(e^{i \delta_m} \xi_{k+m} + e^{-i \delta_m} \xi_{k-m})
\frac{\partial \psi^{(n)}}{\partial {\cal Z}_k}\right|_{\beta_1=\beta_2=...=0}.
\label{difbeta}
\end{equation}
To calculate
the derivative $\partial \psi^{(n)}/ \partial {\cal Z}_k$,
we notice that each of the multivortices can be written as a rational
function,
\begin{mathletters}
\label{MN}
\begin{equation}
\psi^{(n)}=  \frac
{{\cal N}^{(n)}}{{\cal M}^{(n)}},
\end{equation}
where the  numerator
and denominator are homogeneous
 polynomials  in ${\cal Z}_{1-n},{\cal Z}_{2-n},...,{\cal Z}_{n}$,
of degree $P$ ($P \le 2^{n-1}$).
 These  polynomials have the form
\begin{eqnarray}
{\cal N}^{(n)}= {\displaystyle
\sum_{1-n \le i_1, i_2,...i_P \le n}} C_{i_1 i_2...i_P} \,
\delta_{i_1+i_2+...+i_P,n} \,
{\cal Z}_{i_1} {\cal Z}_{i_2}...{\cal Z}_{i_P},
\label{Nn} \\
{\cal M}^{(n)}= {\displaystyle
\sum_{1-n \le i_1, i_2,...i_P \le n-1}} D_{i_1 i_2...i_P} \,
\delta_{i_1+i_2+...+i_P,0}  \,
{\cal Z}_{i_1} {\cal Z}_{i_2}...{\cal Z}_{i_P},
\label{Mn}
\end{eqnarray}
\end{mathletters}where
$C_{i_1 i_2...i_P}$ and $D_{i_1 i_2...i_P}$ are real coefficients
and $\delta_{s,l}$ is Kronecker's delta.
Equations (\ref{MN}) are straightforward from  the recurrence
relation (\ref{v2}).
Note that in eq.(\ref{Nn}),
 products  ${\cal Z}_{i_1} {\cal Z}_{i_2}...{\cal Z}_{i_P}$  have
their indices summing up to $n$; we will be referring to this
property by saying that the polynomial ${\cal N}^{(n)}$ has level $n$.
In this sense, the polynomial ${\cal M}^{(n)}$ (eq.(\ref{Mn}))
has level $0$. When a polynomial of level $l$ is differentiated
w.r.t. ${\cal Z}_k$, its level is lowered down to $l-k$. Therefore,
the derivative $\partial \psi^{(n)}/ \partial {\cal Z}_k$
is a rational function whose numerator is a polynomial
of level $n-k$ and denominator is a polynomial of level $0$.
Since  $\left. {\cal Z}_s \right|_{\beta_1=\beta_2=...=0}= \xi_s
=I_s(r)e^{is \theta}$,
one can easily check that
\[
\left.
\frac{\partial \psi^{(n)}}{\partial {\cal Z}_k}
\right|_{\beta_1=\beta_2=...=0}=
{\frak g}^{(n)}_k(r) e^{i(n-k)\theta}
\]
with some real function ${\frak g}^{(n)}_k(r)$.
Substituting into  (\ref{difbeta}) we obtain
 the zero-frequency eigenfunctions of the operator (\ref{lcsG})
pertaining to the orbital quantum numbers $m$ and $-m$, respectively:
\begin{equation}
\phi_m(r)=\sum_{k=1-n}^n{\frak g}^{(n)}_k(r) I_{k+m}(r);
\quad
\phi_{-m}(r)=  \sum_{k=1-n}^n{\frak g}^{(n)}_k(r) I_{k-m}(r)
\label{phibeta}
\end{equation}
$(m \ge 1)$.  The zero modes (\ref{phibeta}) translate into  zero-frequency
eigenvectors of the operator (\ref{LsG}):
\begin{equation}
\left( \begin{array}{c}
u_m \\ v_m
\end{array} \right) = \sum_{k=1-n}^n {\frak g}^{(n)}_k(r)
\left( \begin{array}{c}
I_{k+m}(r) + I_{k-m}(r) \\ I_{k+m}(r) - I_{k-m}(r)
\end{array} \right).
\label{uvI}
\end{equation}

\section{Concluding remarks}
In this paper we have constructed families of explicit solutions
of the complex sine-Gordon equation.
In literature, the complex sine-Gordon model is usually considered
in the (1+1)-dimensional Minkowski space where it has the form
 \begin{equation}
 \psi_{xx}-\psi_{tt} +
 \frac{(\psi_x^2-\psi_t^2) \, \overline{\psi}}{1-|\psi|^2}
 + \psi(1-|\psi|^2)=0.
 \label{sg_Minkowski}
 \end{equation}
This equation was  introduced in the late 1970s
in three independent field-theoretic
contexts:
(i) as a reduction of the $O(4)$ nonlinear
 $\sigma$-model \cite{Pohlmeyer}; (ii)
 in the description of relativistic strings in a uniform antisymmetric
 tensor field \cite{Lund_Regge}, and (iii) in the theory of
 massless fermions with a scalar contact interaction
 \cite{Neveu_Papanicolaou}. Later on, eq.(\ref{sg_Minkowski}) reappeared
   in an entirely unrelated physical
   context  --- it  turned out to be equivalent to
 the Maxwell-Bloch and self-induced transparency
 equations as well as the system governing stimulated Raman scattering
 in nonlinear optics \cite{Steudel}.
In mathematics literature it is
 common to call eq.(\ref{sg_Minkowski})
  the Lund-Regge model and write it as
   \begin{eqnarray}
  \alpha_{xx}- \alpha_{tt} - \frac{\sin \alpha}{\cos^3 \alpha}
  (\chi_x^2- \chi_t^2) + \sin \alpha \cos \alpha=0,
  \nonumber \\
  (\chi_x \tan^2 \alpha)_x= (\chi_t \tan^2 \alpha)_t,
 \label{Lund-Regge}
 \end{eqnarray}
 where $\sin \alpha$ and $\chi$ are the modulus and argument of the
 complex function $\psi(x,t)$ in (\ref{sg_Minkowski}):
 $\psi = \sin \alpha e^{i \chi}$.
 Geometrically, the Lund-Regge model gives the Gauss-Codazzi
  equations for the embedding of pseudospherical surfaces into
  a flat three-dimensional Euclidean space \cite{Lund_geometry}.

More recent studies of the Lorentzian complex sine-Gordon
theory (\ref{sg_Minkowski}) and its Euclidean counterpart
given by eq.(\ref{sg_Euclidean}),
   were motivated by the fact that
 eqs.(\ref{sg_Minkowski}) and (\ref{sg_Euclidean}) define integrable
 deformations  of 2D conformal field theories,
 more specifically of the $SU(2)/U(1)$ coset model
 and $Z_n$ parafermions \cite{conformal}.
 Other (not unrelated)  sources of interest  have
 been the search for
 exactly solvable conformal theories with black-hole background
 metrics \cite{Saveliev} and
 exact factorisable $S$-matrices
 on the quantum level \cite{quantum}.
  In the current mathematics literature, the
 hyperbolic Lund-Regge equation (\ref{Lund-Regge}) is
 being discussed in connection with the localised induction
 hierarchy describing the motion
 of vortex filaments in an inviscid incompressible fluid
 \cite{Fukumoto}. Its applications to pseudospherical
 surfaces continue to attract attention (see e.g. \cite{Reyes})
 while the elliptic equation
 (\ref{sg_Euclidean})
 has been derived in the description of the ``middle surfaces"
 of generalised Weingarten surfaces \cite{Schief}.

 The (1+1)-dimensional system (\ref{sg_Minkowski})
   was shown to be  integrable
via the inverse scattering transform
 \cite{Pohlmeyer,Lund,G1} and, by virtue of this approach,
 broad classes of its exact solutions have become available
 \cite{G2,dVM2,BG1,BG2}.
The  inverse scattering framework for the
(2+0)-dimensional solitons was formulated
in \cite{BG1}. Solutions obtained in this way
describe nonlinear superpositions of  one-dimensional fronts
intersecting at arbitrary angles on the $(x,y)$-plane \cite{BG1,BG2}.

In Ref.\cite{BP} axially symmetric solutions were constructed
describing $n$ vortices ($n \ge 1$) sitting on top of each other.
In the present paper we have shown that each of these coaxial
multivortices belongs to an infinite-parameter family of
nonsymmetric solutions, which includes, in particular,
configurations of
$n$ single vortices located at separate points of the plane.
(The constellation of $n$ separated monovortices is not the
only splitting possibility for the coaxial $n$-vortex though;
there can also be various combinations of vortices
and antivortices with indices summing up to $n$, the total
topological charge characterising that particular family.)
Our current construction, as well as the one of Ref.\cite{BP},
is based on the B\"acklund transformation of the complex
sine-Gordon equation. (In the axially-symmetric case there
is also an alternative technique employing the Painlev\'e
reduction \cite{BP}.)
For the understanding of the structure of the phase
space, however, it would be instructive to embed the vortex solutions
in the inverse scattering formalism;
we are planning to return to this issue in future publications.
 The fact that even a one-vortex solution is characterised by
an infinite number of parameters seems to indicate
that vortices should be associated with the continuous spectrum
rather than discrete eigenvalues (which account for front-like solitons.)

We have computed the energy (more precisely, the Euclidean action) for each
family of multivortices. The action is found to be entirely determined
 by the total topological charge 
and independent of parameters specifying a particular solution 
within each topological class. This is not an unexpected result;
the nontrivial dependence of the action on some of the continuous
parameters entering the solution
 would contradict the stationarity of this configuration.

That axially-symmetric multivortices of the 
complex sine-Gordon theory admit noncoaxial continuations,
follows already from their spectra of linearised perturbations. As we
have shown in this work, the existence of a zero mode with any
azimuthal number $m=0,1,2,3...$ is ensured by the availability
of the necessary
number of bounded asymptotic solutions
of the linearised equations as $r \to 0$ and $r \to \infty$.
On the contrary, the axially symmetric solutions of the Ginzburg-Landau 
model do not have enough bounded linearised
perturbations as $r \to \infty$; consequently, the existence of 
a zero mode for a particular choice of $n$ and $m$ cannot be 
 guranteed {\it a priori\/}. Our numerical analysis has demonstrated that
the Ginzburg-Landau coaxial multivortices  
 have only
translational and rotational
 zero modes and therefore do not admit continuous deformations and/or
splitting into separate vortices.

\acknowledgements
We are grateful Nora Alexeeva for her help with the  numerics.
Many thanks to Stavros Komineas and Nikos Papanicolaou for
useful discussions on various aspects of topological solitons.
This research was supported by grants from the NRF of South Africa and
the URC of UCT.

\end{document}